\documentclass[a4paper,11pt]{article}
\pdfoutput=1 

\usepackage{jcappub} 
\usepackage{subfigure}
\usepackage{graphicx}
\usepackage{mathtools}
\usepackage{amsmath,bm}
\usepackage[T1]{fontenc} 

\title{\boldmath  Open EFT treatment  of Inflation with Thermal Initial Conditions}

\author[a]{Abbas Tinwala,}
\author[b]{Ashish Narang,}
\author[c]{Subhendra Mohanty,}
\author[a]{Sukanta Panda,}

\affiliation[a]{Indian Institute of Science Education and Research, Bhopal, India}
\affiliation[b]{Institute of Physics, Bhubaneswar, 751005, India}
\affiliation[c]{Indian Institute of Technology, Kanpur, 208016 India}

\emailAdd{abbas18@iiserb.ac.in}
\emailAdd{ashish.narang@iopb.res.in}
\emailAdd{mohantys@iitk.ac.in}
\emailAdd{sukanta@iiserb.ac.in}

\begin{document}
\newcommand{\mb}[1]{\mathbf{#1}}
\newcommand{\bfd}[2]{\mathbf{#1}\cdot\mathbf{#2}}
\newcommand{\ekx}{e^{i\mathbf{k}\cdot\mathbf{x}}}
\newcommand{\mekx}{e^{-i\mathbf{k}\cdot\mathbf{x}}}
\newcommand{\ekxp}{e^{i\mathbf{k}\cdot\mathbf{x'}}}
\newcommand{\mekxp}{e^{-i\mathbf{k}\cdot\mathbf{x'}}}
\newcommand{\eqx}{e^{i\mathbf{q}\cdot\mathbf{x}}}
\newcommand{\meqx}{e^{-i\mathbf{q}\cdot\mathbf{x}}}
\newcommand{\eqxp}{e^{i\mathbf{q}\cdot\mathbf{x'}}}
\newcommand{\meqxp}{e^{-i\mathbf{q}\cdot\mathbf{x'}}}
\newcommand{\ekpx}{e^{i\mathbf{k'}\cdot\mathbf{x}}}
\newcommand{\mekpx}{e^{-i\mathbf{k'}\cdot\mathbf{x}}}
\newcommand{\ekpxp}{e^{i\mathbf{k'}\cdot\mathbf{x'}}}
\newcommand{\mekpxp}{e^{-i\mathbf{k'}\cdot\mathbf{x'}}}
\newcommand{\eqpx}{e^{i\mathbf{q'}\cdot\mathbf{x}}}
\newcommand{\meqpx}{e^{-i\mathbf{q'}\cdot\mathbf{x}}}
\newcommand{\eqpxp}{e^{i\mathbf{q'}\cdot\mathbf{x'}}}
\newcommand{\meqpxp}{e^{-i\mathbf{q'}\cdot\mathbf{x'}}}
\newcommand{\bs}[1]{\boldsymbol{#1}}
\newcommand{\Lopr}{\overrightarrow{\mathbf{\Lambda}}}
\newcommand{\Lopl}{\overleftarrow{\mathbf{\Lambda}}}

\abstract{
     Investigating the thermal inflationary model, we introduce stochastic effects, incorporating a cutoff parameter $\sigma$ which distinguishes between quantum and classical modes. Testing the model against Planck 2018 data, we observe a preference for a non-zero $\sigma$ at least at 68\% C.L., suggesting the classicalization of most modes and providing a theoretical foundation for the quantum to classical transition. As a result of introducing the stochastic effects, we find that the solution to the large-scale power deficit requires a lower comoving temperature of inflaton.}

\maketitle
\section{Introduction}
According to the inflationary paradigm, the large-scale structure observed in the Universe today originates from tiny quantum fluctuations in the early Universe. The inflationary theory suggests that the Universe underwent a rapid expansion phase which resulted in stretching the quantum fluctuations to super-horizon scales. The quantum perturbations of the inflaton field become `classical' at some point in the course of horizon crossing and the classical fluctuations act as seeds for the formation of large-scale structures in the Universe. Over time, under the influence of gravity, these density fluctuations lead to the formation of galaxies and other large-scale structures we observe today ~\cite{Mukhanov:1981xt, Starobinsky:1982ee, Guth:1982ec, Bardeen:1983qw, Mukhanov:1990me}. Detailed observations of cosmic microwave background anisotropy spectrum ~\cite{COBE:1992syq, Planck:2018jri, Hinshaw_2013} and large-scale-structures \cite{Eisenstein:1997ik} provide strong evidence supporting the existence of these super-horizon fluctuations.

One fundamental issue in the study of cosmological density perturbations is an understanding of the exact mechanism which makes the quantum fluctuations in the early universe become classical statistical fluctuations in density which then seed the CMB anisotropy and large-scale structures \cite{Brandenberger:1990bx, Polarski:1995jg, Kiefer:1998qe, Lombardo:2005iz, Burgess:2006jn, Martineau:2006ki, Kiefer:2008ku, Burgess:2014eoa, Burgess:2022nwu}. On the observational front, Bell-type correlations have been proposed with the aim of proving that the observed density perturbations of the CMB and large-scale structures have a quantum origin \cite{Campo:2005sv, Maldacena:2015bha, Martin:2015qta, Martin:2017zxs, Martin:2021znx}.

The stochastic approach provides a framework for converting the effects of quantum fluctuations on the statistical fluctuations in the classical dynamics of the inflaton field.
Stochastic inflation provides a formalism to incorporate quantum fluctuations by introducing a cutoff scale, known as the coarse-graining scale, which distinguishes between short and long-wavelength modes of the initial fluctuations. This formalism enables us to transform a quantum correlator into a statistical one, enabling a more tractable analysis. 
Pioneered by Starobinsky~\cite{Starobinsky:1986fx}, the stochastic formalism was later used to study the effect of random fluctuations in several inflationary set-up~\cite{Nambu:1987ef, Nambu:1988je, Kandrup:1988sc, Nakao:1988yi, Nambu:1989uf, Mollerach:1990zf, Linde:1993xx, Starobinsky:1994bd}. Relatively recently, the stochastic formalism in inflation has been employed to study the generation of primordial black holes~\cite{Vennin:2020kng, Figueroa:2021zah, Ando:2018qdb}, gravitational waves background~\cite{Tasinato:2022asj}, low-$\ell$~\cite{Liguori:2004fa, Wu:2006xp} anomaly, stochastic dark energy~\cite{Glavan:2017jye} and several other scenarios~\cite{Matarrese:2003ye, Geshnizjani:2004tf, Martin:2005ir, Kunze:2006tu, PerreaultLevasseur:2014ziv, Liu:2014ifa, Hardwick:2019uex, Saitou:2019jez, Rigopoulos:2016oko, Tokuda:2018eqs, Mahbub:2022osb, Honda:2023unh}.

The effect of sub-horizon modes on the correlations of super-horizon modes can be computed using path-integrals in the Open EFT formalism \cite{Cheung:2007st, Weinberg:2008hq, Feynman:1963fq, Burgess:2014eoa, Boyanovsky:2015jen, Burgess:2022rdo, Burgess:2022nwu, Colas:2024xjy, Bhattacharyya:2024duw}.
This formalism is useful in the computation of higher-order correlations as in the calculation of non-Gaussianities in the density perturbation during inflation. The Open EFT method 
is useful for generalizing to cases where the initial state of the inflaton may not be the Bunch-Davies vacuum, for example at the start of inflation the ground state of the inflaton could be a thermal state or a squeezed coherent state \cite{Lesgourgues:1996jc, LopezNacir:2011kk, Ghosh:2022cny}. The motivation for studying the quantum-to-classical transition of inflaton perturbations for thermal initial states may also be useful in explaining some existing tension between theory and observation.


The standard slow-roll inflationary model offers a robust explanation for the remarkable uniformity and isotropy of the cosmic microwave background radiation, as well as the observed distribution of galaxies on large scales. It provides a well-established framework for understanding the origins of the large-scale structure in the Universe from quantum fluctuations during the inflationary epoch~\cite{Mukhanov:1981xt, Starobinsky:1982ee, Guth:1982ec, Bardeen:1983qw, Mukhanov:1990me}. However, it falls short in accounting for the significant deficiency in the large-scale power of the Cosmic Microwave Background (CMB) temperature-temperature (TT) mode spectrum, which has been consistently detected through observations by the WMAP~\cite{WMAP:2003elm, Komatsu_2011, Hinshaw_2013} and Planck~\cite{Planck:2018jri} satellites. This inconsistency has been corroborated by the continuously advancing reliability of data. There have been several attempts to resolve the low-$\ell$ problem, including topological defects of cosmic strings in pre-slow-roll era~\cite{Bouhmadi-Lopez:2012izk}, a contracting phase in pre-slow-roll era~\cite{Liu:2013kea}, just-enough inflation~\cite{2012PhRvD..85j3517R}, punctuated inflation~\cite{Jain:2008dw}, cosmic variance~\cite{Iqbal:2015tta}, and resonant superstring excitations during inflation~\cite{Gangopadhyay:2017vqi}.
An inflationary scenario with a pre-inflationary radiation era is one such set-up that provides the explanation for the initial state of the inflation while simultaneously resolving the low-$\ell$ anomaly~\cite{Bhattacharya:2005wn, Powell:2006yg,Wang:2007ws,Das:2014ffa,Cai:2015nya, Bahrami:2015bva, Wang:2016wio, Sasaki:2018ang, Li:2019dnq, Das:2020iah, Anderson:2020hgg}.





In this work, we study a generalization of the stochastic coarse-graining by considering inflation with a preceding thermal era deciding the initial conditions of the inflation while taking into account the stochastic effects on the superhorizon perturbation from the sub-horizon perturbations. We compute the power spectrum using the Schwinger-Keldysh formalism \cite{Schwinger:1960qe, Keldysh:1964ud}. In this technique, we integrate out the sub-horizon modes in the path integrals using a closed time path to account for the thermal initial conditions. This enables us to compute the two-point correlation functions of inflaton taking into account the dissipative effects of sub-horizon modes and the non-vacuum initial states. Our model where we assume that inflation was preceded by a thermal era has two free parameters. These are (1)  the length scale demarcating the separation between the observed super-horizon modes and the integrated out `environmental modes'  and (2) the temperature of the universe at the start of inflation (which commences when the inflaton potential becomes larger than the radiation density). We compare the two-point correlators with these two free parameters with experiments to see if the model provides a better fit than the slow-roll model in the low-$\ell$ TT-anisotropy spectrum of the CMB.

 The paper is organized as follows. In section~\ref{sec:formalism} we briefly recall the thermal inflation and explain the stochastic inflation approach following the influence functional approach. In section~\ref{sec:noise} we calculate the noise correlator. Section~\ref{sec:PPS} provides the primordial power spectrum taking into account the stochastic effects in our inflationary set-up of a stochastic inflation with a pre-inflationary radiation era. In section~\ref{sec:analysis} we provide the details of Monte-Carlo analysis to constrain the parameters of the stochastic inflation and we conclude with results in section~\ref{sec:results}

\section{Thermal initial conditions of the inflation era}

In this section, we recall the thermal inflation following Refs.~\cite{Bhattacharya:2005wn,Powell:2006yg,Wang:2007ws,Das:2014ffa,Cai:2015nya,Bahrami:2015bva,Wang:2016wio,Bhattacharya:2006dm,Sasaki:2018ang,Li:2019dnq,Das:2020iah,Anderson:2020hgg}
and then go on to describe the setup for stochastic inflation with a pre-inflationary radiation era. In this scenario, the universe starts in a radiation era with a slow roll sub-dominant inflaton potential $V(\varphi)$. The inflationary era commences when the temperature falls to a value $T$ where the radiation energy density obeys the relation  $\rho_r = (\pi^2 /30) g_* T^4 < V(\varphi)$. 
The equation governing the evolution of scalar inflaton perturbations $ \phi$ in the radiation era  is given by \cite{Powell:2006yg},
\begin{align}
   {\phi}^{\prime \prime}+2 {\cal H} {\phi}^\prime+V_{,\varphi \varphi} {a^2} \phi + k^2 \phi = 0,
   \label{delta-rad}
\end{align}
where primes denote derivatives w.r.t conformal time $\eta$. In the radiation era, the scale factor goes as $a(\eta)\propto \eta$. Defining $\sigma(\eta)\equiv a(\eta) \phi(\eta)$ the equation (\ref{delta-rad}) in terms of $\sigma$ reduces to
\begin{align}
   \sigma^{\prime \prime} +V_{,\varphi \varphi}  \sigma + k^2\sigma= 0.
   \label{sigma-rad}
\end{align}
Assuming that in the pre-inflation radiation era the inflaton potential is flat and we can ignore the $V_{,\varphi \varphi}$ term \cite{Powell:2006yg,Das:2020iah}, solution of (\ref{sigma-rad} ) which matches with the positive frequency plane waves at large $k$ and obeys the Bunch-Davies initial conditions has the form
\begin{equation}
\sigma_1(\eta)=\frac{1}{\sqrt{2k}} \,e^{-i k \eta}   \quad \quad a(\eta)<a_i , 
\label{sigma-rad1}
\end{equation}
where $a_i$ is the scale factor at the transition from the radiation to the inflation era.

The radiation era is followed by inflation where the conformal time is related to the scale factor as $\eta=-1/(1-\epsilon)a H$. The inflaton perturbations $\sigma=a \phi$ obeys the equation \cite{Powell:2006yg, Das:2020iah},
\begin{align}
   \sigma^{\prime \prime} +\left(k^2- \frac{1}{\eta^2}\left(\nu^2-\frac{1}{4}\right)\right)\sigma= 0, 
   \label{sigma-infl}
\end{align}
where $\nu=(9/4-\eta_V+\epsilon)$. This has the general solution
\begin{equation}
\sigma_2(\eta)=\sqrt{-k \eta}\left( C_1(k)\,H_\nu^{(1)}(-k\eta) + C_2(k)\, H_\nu^{(2)}(-k \eta)\right),    \quad \quad a(\eta)\geq a_i , 
\label{sigma-infl1}
\end{equation}
where $H_\nu^{(1)}(-k\eta)$  and $H_\nu^{(2)}(-k\eta)$ are Hankel functions and where $C_1(k)$ and $C_2(k)$ are integration constants which are determined by matching the solutions (\ref{sigma-rad1}) and (\ref{sigma-infl1}) at the transition $a_i$.  Its useful to take the asymptotic forms of  $H_\nu^{(1)}(-k\eta) $ and $H_\nu^{(1)}(-k\eta) $ in the limit $-k \eta \gg  1 $ to write the solution (\ref{sigma-infl1}) in the form \cite{Powell:2006yg},
\begin{equation}
\tilde \sigma_2(\eta)\equiv   \sigma_2(\eta)\big\vert_{(-k\eta) \gg 1}= \sqrt{-k \eta}\,\, Y_\nu (-k\eta)\left(C_1(k)\, -  C_2(k)\,\right)    
\label{sigma-infl1}
\end{equation}
which is the inflaton perturbation in the inflation era near the transition point and where $Y_\nu (-k\eta)$ is the Bessel function of the second kind. Since the conformal time is not defined continuously across the boundary we use the scale factor as the parameter for imposing the condition that the solutions in the radiation and inflation era and their derivatives w.r.t the scale factor match at $a=a_i$.

In the presence of a pre-inflationary radiation era, the mode solutions are obtained as described in \cite{Das:2014ffa} and are given as,
\begin{equation}
    \phi_\mathbf{k}(x) =\dfrac{1}{(2\pi)^{3/2}} \frac{H\sqrt{\pi}}{2}|\eta|^{3/2}(H_\nu^{(1)}(k|\eta|)C_1(k) + H_\nu^{(2)}(k|\eta|)C_2(k)),\label{genmodes1}
\end{equation}
(the $e^{i\mathbf{k}\cdot\mathbf{x}}$ factor have been taken out from the mode functions in the definition of Fourier transform) where the coefficients $C_1(k)$ and $C_2(k)$ are determined by matching the modes and their derivatives at the transition from the pre-inflationary radiation era to the inflation era. If there is no pre-inflationary radiation era, the coefficients simplify to $C_1(k) = 1$ and $C_2(k) = 0$. The transition from the radiation to the inflation era imposes boundary conditions, leading to an inflationary solution that deviates from the canonical Bunch-Davies solution. The resulting power spectrum, denoted as $\mathcal{P}_{\mathcal R}(k)$, is given by

\begin{eqnarray}
\mathcal{P}_{\mathcal R}(k) = \mathcal{P}_{\mathcal R}^{\rm BD}|C_1 - C_2|^2 = A\left(\frac {k}{k_P}\right)^{n_s - 1}|C_1 - C_2|^2.
\label{powspec1}
\end{eqnarray}

Here, $A$ and $n_s$ represent the amplitude and scalar spectral index, respectively, in a generic inflationary scenario, while $k_P$ denotes the pivot scale. The quantity $|C_1 - C_2|^2$ captures the modification due to mode function matching and is given by,
\begin{eqnarray}
|C_1-C_2|^2&=&\frac{\pi}{2z_I}\left[z_I^2\left(J_{\nu_\chi}^2(z_I)+J_{\nu_\chi+1}^2(z_I)\right)-2z_I\left(\nu_\chi+\frac12\right)J_{\nu_\chi}(z_I)J_{\nu_\chi+1}(z_I)\right.\nonumber\\
&&\left.\left.+\left(\nu_\chi+\frac12\right)^2J_{\nu_\chi}^2(z_I) \right]\right|_{a_i},
\label{factor}
\end{eqnarray}
where $z_I(k,t_i)=k/(1-\epsilon_I)a_iH_i$, where $a_iH_i$ corresponds to the horizon size at the onset of inflation. The largest mode of interest with wavenumber $k_{0}=a_0 H_0=H_0$ and $N(k_{0})-N(k_{i})=\ln(k_i/k_{0})$ leaves $N(k_0)$ e-foldings before the end of inflation. This leads to $k_i=k_{0}e^{-\delta N}$, where $\delta N \equiv N(k_{i})-N(k_{0})$. Finally, $z_I(k,t_i)$ is evaluated as $k/((1-\epsilon_I)H_0)e^{\delta N}$.

Considering a thermal distribution of scalar field quanta, an additional multiplicative factor is introduced in the power spectrum~\cite{Bhattacharya:2005wn}. In this case, the power spectrum is given by
\begin{eqnarray}
\mathcal{P}_{\mathcal R}(k) = A_s^\prime\left(\frac {k}{k_P}\right)^{n_s-1}|C_1-C_2|^2\coth\left(\frac{k}{2T}\right),
\label{pi_pps1}
\end{eqnarray}
where $T$ represents the comoving temperature of the scalar field quanta. In the following sections, we will determine the corrections to this expression for power spectrum due to the dissipative effect of the sub-horizon perturbations which are integrated out to give the influence functional which gives the two-point correlations of the super-horizon modes.

\section{Stochastic Thermal Inflation Formalism}\label{sec:formalism}
Stochastic inflation \cite{Starobinsky:1986fx} is a framework that incorporates the effects of quantum fluctuations during inflation by treating them as stochastic noise. Conventionally, in the stochastic inflation formalism, the inflaton field is divided into a super-horizon part and a sub-horizon part directly within the equation of motion. This splitting is performed in Fourier space using a window function that effectively separates long-wavelength modes from short-wavelength modes. The primary focus is on the long-wavelength component, while the sub-horizon modes are incorporated as an effective noise term, representing classical perturbations that influence the dynamics of the super-horizon field.
An alternative and more general approach utilizes the influence functional method \cite{Morikawa:1989xz, Polarski:1995jg}, which implements the frequency splitting at the action level by integrating out the short-wavelength modes through a path integral over the sub-horizon part of the field. 


The main purpose of this section is to obtain an effective action by integrating the short-wavelength modes of the scalar field in the path integral formalism. The split of the scalar field is done through the use of a window function whose role is to extract the short-wavelength modes while eliminating the long ones from the scalar field. As such we can obtain the small-wavelength modes from the scalar field in the following way,
\begin{align}\label{shortphi}
    \phi_>(x) = \int dk W(k,t) \phi_{k}(x).
\end{align}
The convention of using the `greater-than sign' for the short-wavelength modes is common in literature as it is a common practice to distinguish modes based on their wavenumbers rather than their wavelengths. The exact form of the window function has been a subject of discussion in past work. The simplest one is the Heaviside theta function $\theta(k-\sigma aH)$ \cite{10.1007/3-540-16452-9_6}. The calculations are simplest with this choice since it leads to white noise. However, this choice is ideal in the sense that it abruptly cuts off all the wavelengths less than $\sigma aH$ from $\phi_<$. Therefore it makes more sense to have a Window function that smoothly decays from 1 to 0 as we increase the wavelength of modes beyond the cutoff. This leads to a general class of exponential filters\cite{Mahbub:2022osb},
\begin{align}
    W(k\eta) = 1-\text{exp}\left\{-\dfrac{1}{2}\left(\dfrac{k\eta}{\sigma}\right)^n\right\}.
\end{align}
We will use $n=2$ in this work following references \cite{Winitzki:1999ve,Matarrese:2003ye} which is a Gaussian filter.
We split the scalar field into short-wavelength and long-wavelength parts in the total action $S[\phi]$ of the scalar field $\phi$ to obtain
\begin{align}\label{Sint}
    S[\phi]=S[\phi_<] + S[\phi_>] + S_{\text{int}}[\phi_<,\phi_>],
\end{align}
where $S_{\text{int}}$ is the interaction between long and short-wavelength parts of the scalar field arising as a result of making the split. Note that had we started with a free theory of a scalar field we would still obtain a coupling between the long and short-wavelength parts. The interaction term reads,
\begin{align}
    S_{\text{int}}=\int d^4 x \phi_> (x) \Lambda(x) \phi_< (x),
\end{align}
where
\begin{align}\label{Lambda}
    \Lambda(x)= -a^3(t)\left(\partial_t^2+3H\partial_t-\dfrac{\nabla^2}{a^2(t)}+m^2\right).
\end{align}
Since we are interested in obtaining the power spectrum of the long-wavelength modes which is an \textit{in-in} expectation value we will employ the \textit{Closed Time Path} (CTP) formalism.

The \textit{CTP} formalism \cite{Schwinger:1960qe} is used to compute the quantum averages of operators by specifying only the \textit{in}-state without requiring to specify the \textit{out}-state. This makes it different from the usual quantum field theory which is used to compute the quantum averages of operators by specifying the \textit{in}-state and the \textit{out}-state. The time evolution of the state follows a closed-time path in the \textit{CTP} formalism where the system evolves from an initial time where the \textit{in}-state is specified, forward to $t=\infty$ and then back again to the initial time. This forward-backward evolution is avoided in the usual quantum field theory where the system is assumed to undergo adiabatic evolution from its ground state. The quantum averages of the operators are computed on this forward-backward path through path integral formalism where the field configuration does not necessarily assume the same values on the forward and backward branches. As such we double the degrees of freedom by considering two fields $\phi^+$ and $\phi^-$ residing on the forward and backward branches of the time contour, respectively, with the constraint $\phi^+(\infty)=\phi^-(\infty)$. The whole point of the CTP formalism is that for out-of-equilibrium systems the fields on the forward and backward are not the same to begin with and are matched only after having derived the equations of motion. Additionally, we have also made a split based on wavelength and so we end up with four fields namely, $\phi_>^+$, $\phi_>^-$, $\phi_<^+$ and $\phi_<^-$. To obtain the effective action $\Gamma[\phi_<]$ we path-integrate the action over $\phi_>^+$ and $\phi_>^-$, 
\begin{align}\label{effac}
    e^{i\Gamma[\phi_<^\pm]} = e^{i(S[\phi_<^+]-S[\phi_<^-])}\int \mathcal{D}\phi_>^\pm e^{i(S[\phi_>^+]-S[\phi_>^-])+i(S_{\text{int}}[\phi_<^+,\phi_>^+]-S_{\text{int}}[\phi_<^-,\phi_>^-])}.
\end{align}
A straightforward calculation results in the following expression for the effective action (see Ref.~\cite{Morikawa:1989xz,Matarrese:2003ye, PhysRevD.88.083537} for details)
\begin{align}
    \Gamma[\varphi^\pm]=S[\varphi^+]-S[\varphi^-]+S_{\text{inf}}
\end{align}
where $\varphi^\pm = \phi_<^\pm$. The influence functional $S_{\text{inf}}$ is given by

\begin{align}
    S_{\text{inf}}=-\dfrac{1}{2}\int d^4x d^4x'\boldsymbol{\varphi}^T(x)\Lopr(x) \mathbf{\Lambda}^{-1}(x,x')\Lopl(x')\boldsymbol{\varphi}(x'),\label{Sinf}
\end{align}
where the column vector $\boldsymbol{\varphi}$ is given by
\begin{align}
    \boldsymbol{\varphi} = 
    \begin{pmatrix}
        \varphi^+ \\ \varphi^-
    \end{pmatrix},
\end{align}
and the matrix operator $\mathbf{\Lambda}$ is given by
\begin{align}
\mathbf{\Lambda}=
\begin{pmatrix}
\Lambda & 0\\0 & -\Lambda
\end{pmatrix},
\end{align}
whose inverse is the Green's function on the contour for the field $\phi_>$ given by
\begin{align}\label{propmatrix}
\mathbf{\Lambda}^{-1}(x,x') =-i
\begin{pmatrix}
\langle T[\hat\phi^+_>(x)\hat\phi^+_>(x')]\rangle & \langle \hat\phi^-_>(x')\hat\phi^+_>(x)\rangle \\\langle \hat\phi^-_>(x)\hat\phi^+_>(x')\rangle &\langle \bar T[\hat\phi^-_>(x)\hat\phi^-_>(x')]\rangle
\end{pmatrix}.
\end{align}
The arrows over the matrix operator $\mathbf{\Lambda}$ indicate the direction in which the space and time derivatives it contains are acting. The upper-left element of $\mathbf{\Lambda}^{-1}$ is the usual time-ordered correlator of fields lying on the forward contour. The lower-right element is the anti-time-ordered correlator of fields lying on the backward branch of the counter. The anti-time ordering appears because, on the backward branch of the contour, the time evolution happens from $+\infty$ to $-\infty$ due to which the fields at later times appear `before' those at earlier times (of course, a more appropriate term to use is `path-ordering' rather than `time ordering'). The upper(lower) off-diagonal element is proportional to the negative(positive) frequency Wightman propagator. Path-ordering tells us that these are absolutely-ordered correlators since the field on the forward branch must always appear `before' the field on the backward branch.
\subsection{Thermal Influence Functional}
Let us compute the influence functional Eq.~(\ref{Sinf})  for the case of a massive scalar field whose modes start from an initial thermal state. The Fourier transform of the short-wavelength field is given by,
\begin{align}\label{ftsmall}
    \phi_>(x)&=\int d^3k \ W(k,t)(\phi_{\mathbf{k}}(t)\hat a_{\mathbf{k}} e^{-i\mathbf{k}.\mathbf{x}}+\phi^*_\mathbf{k}(t)\hat a^\dagger_\mathbf{k} e^{i\mathbf{k}.\mathbf{x}})\nonumber\\
    &=\int d^3k \  W(k,t)(\phi_{\mathbf{k}}(t)\hat a_{\mathbf{k}}+\phi^*_{-\mathbf{k}}(t)\hat a^\dagger_{-\mathbf{k}} ) e^{-i\mathbf{k}.\mathbf{x}}.
\end{align}
Using Eqs.~(\ref{ftsmall}) and (\ref{propmatrix}) in Eq.~(\ref{Sinf}), we find,
\begin{align}
S_{\text{inf}}=\left(\dfrac{i}{2}\right)\int d^4x d^4x' \int d^3\mathbf{k} \mekx\bs{\varphi}^T_\mb{k}(t)\int d^3\mb{q}&\meqxp\int d^3\mb{k'}\int d^3\mb{q'}\mekpx\meqpxp\nonumber\\
    &\times\Lopr\left[ W(k',t) 
    \mathbf{\hat Q}(t,t')\right]W(q',t') \Lopl \bs{\varphi}_\mb{q}(t'), \label{qm}
\end{align}
where
\begin{equation} \label{Qmatrix}
\begin{split}
&\mathbf{\hat Q}(t,t')\\
    &=\begin{pmatrix}
        \begin{multlined}
             \langle T[\phi_\mb{k'}(t)\phi^*_\mb{-q'}(t')\hat a_\mb{k'}\hat a^\dagger_\mb{-q'}       \\[-2ex]
            \hspace{23mm}+\phi^*_\mb{-q'}(t')\phi_\mb{k'}(t)\hat a^\dagger_\mb{-q'}\hat a_\mb{k'}]\rangle
        \end{multlined} 
        &\begin{multlined}
            \langle \phi_\mb{q'}(t')\phi^*_\mb{-k'}(t)\hat a_\mb{q'}\hat a^\dagger_\mb{-k'}\\[-2ex]
             \hspace{23mm}+\phi^*_\mb{-q'}(t')\phi_\mb{k'}(t)\hat a^\dagger_\mb{-q'}\hat a_\mb{k'}\rangle
        \end{multlined} 
        \\[4ex]
      \begin{multlined}
             \langle \phi_\mb{k'}(t)\phi^*_\mb{-q'}(t')\hat a_\mb{k'}\hat a^\dagger_\mb{-q'}       \\[-2ex]
            \hspace{23mm} +\phi^*_\mb{-q'}(t')\phi_\mb{k'}(t)\hat a^\dagger_\mb{-q'}\hat a_\mb{k'}\rangle
        \end{multlined} 
        & \begin{multlined}
             \langle \bar T[\phi_\mb{k'}(t)\phi^*_\mb{-q'}(t')\hat a_\mb{k'}\hat a^\dagger_\mb{-q'}       \\[-2ex]
            \hspace{23mm} +\phi^*_\mb{-q'}(t')\phi_\mb{k'}(t)\hat a^\dagger_\mb{-q'}\hat a_\mb{k'}]\rangle
        \end{multlined} 
    \end{pmatrix}.
\end{split}
\end{equation}
Strictly, the exponentials $\mekpx$ and $\meqpxp$ cannot be taken out from the square brackets in Eq.~(\ref{qm}) since the operator $\mathbf{\Lambda}$ also contains space derivatives. However, the space derivatives and the mass terms can be thrown away for reasons that will become clear.
The action of the matrix operator is given by
\begin{align}
    &\Lopr [W(k,t)\mb{\hat{Q}}(t,t')W(k,t')] \Lopl \nonumber\\&\hspace{25mm}=a^3(t)a^3(t')\bs{\sigma}_3\left(\dfrac{\partial'^2}{\partial t'^2}+3H\dfrac{\partial'}{\partial t'}\right)\Big[W(k,t')\dfrac{\partial^2}{\partial t^2}(W(k,t) \mb{\hat{Q}}(t,t')) \nonumber\\&\hspace{40mm}+ 3HW(k,t')\dfrac{\partial}{\partial t}(W(k,t)\mb{\hat{Q}}(t,t')\Big]\bs{\sigma}_3\\
    &\hspace{25mm}= a^3(t)a^3(t')\bs{\sigma}_3 \Big\{\overrightarrow{P}_t\mb{\hat{Q}}(t,t')\overleftarrow{P}_{t'}+\overrightarrow{P}_t\mb{\hat{R}}(t,t')W(k,t')+\mb{\hat{R}}(t,t')\overleftarrow{P}_{t'}W(k,t)\nonumber\\&\hspace{40mm}+\left(\dfrac{\partial^2}{\partial t^2}\mb{\hat R} + 3H\dfrac{\partial}{\partial t}\mb{\hat R}(t,t')\right)W(k,t)W(k,t')\Big\}\bs{\sigma}_3,\label{Lopraction}
\end{align}
where $\bs{\sigma}_3$ is the Pauli matrix $\begin{pmatrix} 1 & 0 \\ 0 & -1 \end{pmatrix}$, and $\mb{\hat R}(t,t')$ is given by
\begin{align}
    \mb{\hat R}(t,t') = \dfrac{\partial^2
}{\partial t^2}\mb{\hat Q}(t,t')+3H\dfrac{\partial}{\partial t}\mb{\hat Q}(t,t').
\end{align}
In the expression (\ref{Lopraction}) we will neglect all those terms that do not involve time derivatives of the window function. The reason behind this is that the role of the window function is to project out the short wavelength modes and since $\varphi_\mathbf{k}$ contains only the long wavelength modes by definition, we expect $W(k,t)\varphi_\mathbf{k}$ to vanish. Now it is clear why we can throw away the spatial derivatives and mass terms in the operator $\mathbf{\Lambda}$ since these give rise to terms that do not involve the time derivative of the window functions.

For inflation following a thermal radiation era, the initial thermal state is defined by a bosonic occupation number instead of a zero particle vacuum state. For bosons, the average occupation number of a state with momentum $\mb{k}$ and temperature $T=1/\beta$ is given by
\begin{align}
    n(\omega_k) = \frac{1}{e^{\beta\omega_k}-1},
\end{align}
where $\omega_k = \sqrt{|\mb{k}|^2+m^2}$.
We find the following expectation values involving the creation and annihilation operators for the initial thermal state,
\begin{align}\label{caexpec}
&\langle \hat a^\dagger_\mb{k}\hat a_\mb{q}\rangle = n(\omega_k)\delta^3(\mb{k}-\mb{q}), \nonumber \\
&\langle \hat a_\mb{k}\hat a^\dagger_\mb{q}\rangle = (1+n(\omega_k))\delta^3(\mb{k}-\mb{q}).
\end{align}
Integrating over $x$ and $x'$ in Eq.~(\ref{qm}), and using Eq.~(\ref{Lopraction}) keeping in mind that at least a single time derivative must be acting on the window function followed by using the expectation values (\ref{caexpec}), we find,
\begin{equation}
\begin{split}
&S_{\text{inf}}=\left(\dfrac{i}{2}\right)\int dt dt' \ a^3(t) \int d^3\mathbf{k}  \ \bs{\varphi}^T_{-\mb{k}}(t)\\
    &\times \overrightarrow{P}_t
    \begin{pmatrix}
            \begin{multlined}
                (\theta(t-t')+n)\phi_\mb{k}(t)\phi^*_\mb{k}(t')     
             \\[-2.5ex]
                \hspace{2mm}+ (\theta(t'-t)+n)\phi^*_\mb{-k}(t)\phi_\mb{k}(t')
            \end{multlined}
        &   \begin{multlined}
               -(1+n)\phi_\mb{-k}(t')\phi^*_\mb{-k}(t)
            \\[-2.5ex]
                \hspace{2mm}-n\phi^*_\mb{k}(t')\phi_\mb{k}(t)
            \end{multlined} 
        \\[4ex]
            \begin{multlined}
                -(1+n)\phi_\mb{k}(t)\phi^*_\mb{k}(t')
            \\[-2.5ex]
                \hspace{2mm}-n\phi^*_\mb{-k}(t)\phi_\mb{-k}(t')
            \end{multlined}
        &  \begin{multlined}
                (\theta(t'-t)+n)\phi_\mb{k}(t)\phi^*_\mb{k}(t')     
             \\[-2.5ex]
                \hspace{2mm}+ (\theta(t-t')+n)\phi^*_\mb{-k}(t)\phi_\mb{k}(t')
            \end{multlined}
    \end{pmatrix}\overleftarrow{P}_{t'} a^3(t') \bs{\varphi}_\mb{k}(t').
\end{split}
\end{equation}\label{eq:afterIFT}
Using the inverse Fourier transform of the field modes $\varphi_\mb{k}$ we find,
\begin{equation}\label{ift}
\begin{split}
&S_{\text{inf}}=\left(\dfrac{i}{2}\right)\int d^4x d^4x' \ a^3(t) \int d^3\mathbf{k}  \ \bs{\varphi}^T(x)\times\\
    & \overrightarrow{P}_t
    \begin{pmatrix}
            \begin{multlined}
                (\theta(t-t')+n)\phi_\mb{k}(t)\phi^*_\mb{k}(t')\mekx\ekxp     
             \\[-2.5ex]
                \hspace{-5mm}+ (\theta(t'-t)+n)\phi^*_\mb{k}(t)\phi_\mb{k}(t')\ekx\mekxp
            \end{multlined}
        &   \begin{multlined}
               -(1+n)\phi_\mb{k}(t')\phi^*_\mb{k}(t)\ekx\mekxp
            \\[-2.5ex]
               \hspace{-5mm}-n\phi^*_\mb{k}(t')\phi_\mb{k}(t)\mekx\ekxp
            \end{multlined} 
        \\[5ex]
            \begin{multlined}
                -(1+n)\phi_\mb{k}(t)\phi^*_\mb{k}(t')\mekx\ekxp
            \\[-2.5ex]
               \hspace{-5mm}-n\phi^*_\mb{k}(t)\phi_\mb{k}(t')\ekx\mekxp
            \end{multlined}
        &   \begin{multlined}
                (\theta(t'-t)+n)\phi_\mb{k}(t)\phi^*_\mb{k}(t')\mekx\ekxp     
             \\[-2.5ex]
                \hspace{-5mm}+ (\theta(t-t')+n)\phi^*_\mb{k}(t)\phi_\mb{k}(t')\ekx\mekxp
            \end{multlined}
    \end{pmatrix}\overleftarrow{P}_{t'}\\
    &\times a^3(t') \bs{\varphi}(x').
\end{split}
\end{equation}
To facilitate taking the limit $\varphi_+ = \varphi_-$ when finding the equations of motion it is convenient to work with the Keldysh basis:
\begin{align}
    \tilde{\bs{\varphi}} = \begin{pmatrix} \varphi_c \\ \varphi_q\end{pmatrix} = \mb{U}\bs{\varphi} = \begin{pmatrix} \dfrac{(\varphi^++\varphi^-)}{2} \\ \varphi^+-\varphi^-\end{pmatrix}, \hspace{15mm}\text{where  } \mb{U}=\begin{pmatrix}  1/2 & 1/2 \\ 1 & -1 \end{pmatrix}
\end{align}
To make the change of basis we insert factors $\mb{U}^T(\mb{U}^T)^{-1}$ and $\mb{U}^{-1}\mb{U}$ at appropriate places as shown below.
\begin{equation}
\left(\dfrac{i}{2}\right)\int d^4x d^4x' \ a^3(t)\underbrace{\bs{\varphi}^T(x) \mb{U}^T}_{=\tilde{\bs{\varphi}}^T(x)}\int d^3\mathbf{k} \ (\mb{U}^T)^{-1} \mb{P} \mb{U}^{-1} \ \underbrace{\mb{U}\bs{\varphi}^(x')}_{=\tilde{\bs{\varphi}}}a^3(t'),
\end{equation}
where the matrix $\mb{P}$ is the product $\overrightarrow{P}_t ( ... )\overleftarrow{P}_{t'}$ in Eq.~(\ref{ift}).
The matrix product $\tilde{\mb{P}}=(\mb{U}^T)^{-1} \mb{P} \mb{U}^{-1}$ is straightforward to evaluate,
\begin{equation}\tilde{\mb{P}}=
    \begin{pmatrix}
         0
        &  \begin{multlined}
               \theta(t'-t)\{\phi_\mb{k}(t')\phi^*_\mb{k}(t)\mekx\ekxp
            \\[-2.5ex]
               -\phi_\mb{k}(t)\phi^*_\mb{k}(t')\mekx\ekxp\}
            \end{multlined} 
        \\[5ex]
            \begin{multlined}
               \theta(t-t')\{\phi_\mb{k}(t)\phi^*_\mb{k}(t')\mekx\ekxp
            \\[-2.5ex]
               -\phi^*_\mb{k}(t)\phi_\mb{k}(t')\mekx\ekxp\}
            \end{multlined}
         &   \begin{multlined}
                \Big(n+\dfrac{1}{2}\Big)\Big(\phi_\mb{k}(t)\phi^*_\mb{k}(t')\mekx\ekxp   
             \\[-2.5ex]
                + \phi^*_\mb{k}(t)\phi_\mb{k}(t')\mekx\ekxp\Big)
            \end{multlined}
    \end{pmatrix}.
\end{equation}
Finally, on expanding the matrix products and doing further simplifications, the influence functional can be written as,
\begin{align}
    S_{\text{inf}} &= \int d^4x d^4x' \Bigg\{\dfrac{i}{\hbar}\left(n+\dfrac{1}{2}\right)\varphi_q(x)\text{Re}[\Pi(x,x')]\varphi_q(x')-\dfrac{2}{\hbar}\varphi_q(x)\theta(t-t')\text{Im}[\Pi(x,x')]\varphi_c(x')\Bigg\}\nonumber\\ &=S_{\text{I}}+S_{\text{R}},\label{eq:Sinf}
\end{align}
where 
\begin{align}
    \Pi(x,x') = \int d^3\mb{k} \ a^3(t)a^3(t')\ [P_t\phi_\mb{k}(t)]\ [P_{t'}\phi^*_\mb{k}(t')]\mekx\ekxp.
\end{align}
We will call $S_{\text{inf}}$ the thermal influence functional as it differs from the one derived earlier in Ref.~\cite{Matarrese:2003ye, Morikawa:1989xz, PhysRevD.88.083537} by the presence of the average occupation number.

Before we proceed to carry out an explicit calculation of the two terms that appear in the expression of the thermal influence functional in the next section we pause a little to say something about the thermal influence functional. We observe that $S_{\text{inf}}$ contains an imaginary part $S_{\text{I}}$ along with the usual real part $S_{\text{R}}$. How do we interpret the imaginary part of the effective action? As we will show in the next section this term can be interpreted as a result of the statistical weighting over the configurations of the random noise fields representing the short-wavelength quantum fluctuations. The noise influences the evolution of the super-horizon modes as we shall see when we derive the equations of motions in section \ref{sec:PPS}. The real part of the thermal influence functional gives rise to dissipation and mass renormalization. The dissipation term is basically a friction term, i.e. proportional to $H\dot\varphi$ as shown in Appendix \ref{app:dissipation}. There we also show that both the dissipation and mass renormalization terms are negligible compared to other terms in the equations of motion for the super-horizon fluctuations as long as $\sigma$ is less than one but sufficiently larger than $k|\eta|$.

\subsection{Noise Correlator}\label{sec:noise}
Let us evaluate further the first term in Eq.~(\ref{eq:Sinf}) for the case of a massive scalar field in an initial thermal state. After integrating by parts, the integral over $\mathbf{k}$ in the imaginary part of the influence functional, Eq.~(\ref{eq:Sinf}), reads
\begin{equation}
    \int d^3\mathbf{k}e^{-i\mathbf{k}\cdot(\mathbf{x}'-\mathbf{x})}\coth{(\beta k/2)} \dot{W}(k\eta)\dot{W}(k\eta')(\dot{\phi_k}(t)\varphi_q(x)-\phi_k(t)\dot{\varphi_q}(x))(\dot{\phi_k}^*(t')\varphi_q^*(x')-\phi_k^*(t')\dot{\varphi_q}^*(x')),    
\end{equation}
where $\coth(\beta k/2) = 2n(\omega_k)+1$.

From Eq.~(\ref{genmodes1}) we find
\begin{equation}
    \dot{\phi_k}(t) = -H(\phi_{1k}q_{1\nu}(k\eta)+\phi_{2k}q_{2\nu}(k\eta)),
\end{equation}
where we have defined
\begin{align}
    &\phi_{1k} = \dfrac{1}{(2\pi)^{3/2}}\frac{H\sqrt{\pi}}{2}|\eta|^{3/2}H_\nu^{(1)}(k|\eta|)C_1(k),\\
    &\phi_{2k}=\dfrac{1}{(2\pi)^{3/2}}\frac{H\sqrt{\pi}}{2}|\eta|^{3/2}H_\nu^{(2)}(k|\eta|)C_2(k),\\
    &q_{1\nu}(k\eta) = \frac{3}{2}-\nu+k|\eta|\frac{H_{\nu-1}^{(1)}}{H_\nu^{(1)}}, \text{    and}\\ 
    &q_{2\nu}(k\eta) = \frac{3}{2}-\nu+k|\eta|\frac{H_{\nu-1}^{(2)}}{H_\nu^{(2)}}
\end{align}
After some manipulations, the imaginary part of the influence functional can be written as,
\begin{align}\label{ImSinf}
S_{\text{I}}=   \frac{i}{2}\int d^4x d^4x'a^3(t)a^3(t')[\psi_q(x),\varphi_q(x)]\mathbf{A}(x,x')\begin{bmatrix} \psi_q(x')\\ \varphi_q(x') \end{bmatrix},
\end{align}
where we have defined
\begin{align}
    &\psi_q(x) = \varphi_q\left(\frac{3}{2}-\nu\right)+\frac{\dot{\varphi_q}}{H},\\
    &\mathbf{A}(x,x')= \int d^3\mathbf{k}e^{-i\mathbf{k}\cdot(\mathbf{x}'-\mathbf{x})}\frac{k^2}{a(t)a(t')}W'(k\eta)W'(k\eta')e^{i\mathbf{k}\cdot(\mathbf{x'}-\mathbf{x})}H^2\mathbf{D}(k\eta,k\eta'),\label{noisecorr}
\end{align}
where
\begin{align}
    &\mathbf{D}_k = \begin{pmatrix} (\phi_{1k} + \phi_{2k})(\phi^*_{1k}+\phi^*_{2k}) & -k\eta'(\phi_{1k}(\eta)+\phi_{2k}(\eta))u_k^*(\eta') \\ -k\eta(\phi^*_{1k}(\eta')+\phi^*_{2k}(\eta'))u_k(\eta) & k^2\eta\eta' u_k(\eta)u_k^*(\eta')\end{pmatrix},\\
    & u_k(\eta) = \frac{H_{\nu-1}^{(1)}(k|\eta|)}{H_\nu^{(1)}(k|\eta|)}\phi_{1k}+\frac{H_{\nu-1}^{(2)}(k|\eta|)}{H_\nu^{(2)}(k|\eta|)}\phi_{2k}.
\end{align}
As shown in \cite{Matarrese:2003ye, Morikawa:1989xz, PhysRevD.88.083537} the imaginary part of the influence functional can be interpreted as the result of the statistical weighting over the configurations of the random noise fields which represent the short-wavelength quantum fluctuations. This can be explicitly shown through the application of the Hubbard-Stratonovich transformation which is the following identity when extended over symmetric operators.
\begin{align}
    \text{exp}\left\{-\dfrac{1}{2}\phi_m A_{mn}\phi_n\right\}=N\int\prod_{i=1}^Nd\xi_i \ \text{exp}\left\{-\dfrac{1}{2}\xi_m(A^{-1})_{mn}\xi_n-i\phi_m\xi_m\right\}.
\end{align}
This identity can be used to rewrite Eq.~(\ref{ImSinf}) as follows,
\begin{align}
    \text{exp}\left\{iS_{\text{I}}\right\} &= \text{exp}\left\{-\dfrac{1}{2}\int d^4x d^4x'a^3(t)a^3(t')[\psi_q(x),\varphi_q(x)]\mathbf{A}(x,x')\begin{bmatrix} \psi_q(x')\\ \varphi_q(x') \end{bmatrix}\right\}\nonumber\\
    &=\int \mathcal{D}\xi_1\mathcal{D}\xi_2\text{exp}\left\{-\dfrac{H^2}{2}\int d^4x d^4x'[\xi_1(x),\xi_2(x)]\mb{A}^{-1}(x,x')\begin{bmatrix} \xi_1(x')\\ \xi_2(x') \end{bmatrix}\right\}\nonumber\\
    &\hspace{20mm}\times \text{exp}\left\{i\int d^4x a^3(t)H[\xi_1(x),\xi_2(x)]\begin{bmatrix} \psi_q(x)\\ \varphi_q(x) \end{bmatrix}\right\}\nonumber\\
    &=\int \mathcal{D}\xi_1\mathcal{D}\xi_2\mathcal{P}[\xi_1,\xi_2]\text{exp}\left\{i\int d^4x a^3(t)H[\xi_1(x),\xi_2(x)]\begin{bmatrix} \psi_q(x)\\ \varphi_q(x) \end{bmatrix}\right\}.
\end{align}
From this, we can write Eq.~(\ref{effac}) as follows.
\begin{align}
    e^{i\Gamma[\varphi^\pm]} = \int \mathcal{D}\xi_1\mathcal{D}\xi_2\mathcal{P}[\xi_1,\xi_2]e^{iS_{\text{eff}}},
\end{align}
where
\begin{align}
    S_{\text{eff}}=S[\varphi^+]-S[\varphi^-]+\int d^4x a^3(t)H(\psi_q(x)\xi_1(x)+\varphi_q(x)\xi_2(x))
\end{align}

The two classical real fields act as random noise in the equations of motion with a probability distribution function given by
\begin{align}
    \mathcal{P}[\xi_1,\xi_2]=\text{exp}\left\{-\dfrac{H^2}{2}\int d^4x d^4x'[\xi_1(x),\xi_2(x)]\mb{A}^{-1}(x,x')\begin{bmatrix} \xi_1(x')\\ \xi_2(x') \end{bmatrix}\right\}.
\end{align}

\section{Power Spectrum}\label{sec:PPS}
We are concerned with the computation of the power spectrum for a massive scalar field where we consider the contribution from the stochastic effects arising from the short-wavelength modes `integrated-out' from the action in the previous section. We also assume that the modes begin from a pre-inflationary thermal era. The equation of motion for the scalar field that follow from $S_{\text{eff}}$ is
\begin{align}\label{eq:eom}
    \dfrac{\delta S_{\text{eff}}}{\delta \varphi_q}\Bigg|_{\varphi_q=0}=\ddot{\phi}+3H\dot{\phi}+m^2\phi - \frac{\nabla^2\phi}{a^2} - 3H\xi_1\left(1-\dfrac{1}{3}\left(\dfrac{3}{2}-\nu\right)\right)-\dot{\xi}_1+H\xi_2=0.
\end{align}
The entire effect of the random force is then treated as a perturbation of the classical dynamics. As such we split the field into the average of the full field plus the fluctuation. The average of the field $\bar\varphi$ satisfies
\begin{align}
    \ddot{\bar\varphi}+3H\dot{\bar\varphi}+m^2\bar\varphi - \frac{\nabla^2\bar\varphi}{a^2}=0,
\end{align}
and the fluctuation $\delta\varphi$ satisfies
\begin{align}    
    \ddot{\delta\phi} +3H\dot{\delta\phi}+m^2\delta\phi - \frac{\nabla^2\delta\phi}{a^2} = 3H\xi_1\left(1-\dfrac{1}{3}\left(\dfrac{3}{2}-\nu\right)\right)+\dot{\xi}_1-H\xi_2.
\end{align}

Since we are interested in the super-horizon power spectrum we neglect the exponentially suppressed term involving spatial gradient and also drop the noise derivative and the second-order time derivative of $\delta\phi$ assuming the slow roll conditions to hold \cite{Matarrese:2003ye}, giving us
\begin{align}
    \dot{\delta\phi} = \xi_1\alpha-\frac{\xi_2}{3}-\frac{m^2}{3H}\delta\phi,
\end{align}
where we defined $\alpha = \left(1-\dfrac{1}{3}\left(\dfrac{3}{2}-\nu\right)\right)$. The two-point correlation for $\delta\phi$ then reads,
\begin{align}
    \langle\delta\phi(t)\delta\phi(t')\rangle = e^{-\frac{m^2}{3H}(t-t_i)}e^{-\frac{m^2}{3H}(t'-t_i)}&\int_{t_i}^t\int_{t_i}^{t'} dt \ dt' \ e^{\frac{m^2}{3H}(t-t_i)}e^{\frac{m^2}{3H}(t'-t_i)} \Big(A_{11}(x,x')\alpha^2\nonumber\\&
    -\frac{A_{12}(x,x')\alpha}{3}-\frac{A_{21}(x,x')\alpha}{3}+\frac{A_{22}(x,x')}{9}\Big),
    \end{align}
    where $A_{ij}(x,x')$ are the matrix elements of $\mathbf{A}(x,x')$ and $t_i$ is the time when inflation started.
    
Substituting the noise correlator Eq.~(\ref{noisecorr}) and using $\frac{m^2}{3H^2} = \eta_V$ and $e^{\frac{m^2}{3H}t} = a^{\frac{m^2}{3H^2}}$ we find
\begin{align}
    &\langle\delta\phi(t)\delta\phi(t')\rangle = \int d(\text{ln}k)\coth{\left(\frac{\beta k}{2}\right)} \ k^{2\eta_V}(aH)^{-\eta_V}(a'H)^{-\eta_V}\frac{H^2}{8\pi\sigma^4}\nonumber\\
    &\hspace{20mm}\times\Big[\int_{k\eta}^{k\eta_{i}}dx\Big(\alpha(-x)^{\frac{5}{2}-\eta_V}e^{-\frac{x^2}{2\sigma^2}}(C_1H_\nu^{(1)}(x)+C_2H_\nu^{(2)}(x))\nonumber\\&\hspace{50mm}-\frac{x}{3}(-x)^{\frac{5}{2}-\eta_V}e^{-\frac{x^2}{2\sigma^2}}(C_1H_{\nu-1}^{(1)}(x)+C_2H_{\nu-1}^{(2)}(x))\Big)\Big]^2\label{phiphi1}.
   \end{align}
The integral over $x$ can be split into super-horizon and sub-horizon integrals respectively as follows.
\begin{align}\label{split}
    &\int_{k\eta}^{1}dx\Bigg(\alpha(-x)^{\frac{5}{2}-\eta_V}e^{-\frac{x^2}{2\sigma^2}}(C_1H_\nu^{(1)}(x)+C_2H_\nu^{(2)}(x))\nonumber\\&\hspace{60mm}-\frac{x}{3}(-x)^{\frac{5}{2}-\eta_V}e^{-\frac{x^2}{2\sigma^2}}(C_1H_{\nu-1}^{(1)}(x)+C_2H_{\nu-1}^{(2)}(x))\Bigg)\nonumber\\&+\int_{1}^{k|\eta_{i}|}dx\Big(\alpha(-x)^{\frac{5}{2}-\eta_V}e^{-\frac{x^2}{2\sigma^2}}(C_1H_\nu^{(1)}(x)+C_2H_\nu^{(2)}(x))\nonumber\\&\hspace{60mm}-\frac{x}{3}(-x)^{\frac{5}{2}-b}e^{-\frac{x^2}{2\sigma^2}}(C_1H_{\nu-1}^{(1)}(x)+C_2H_{\nu-1}^{(2)}(x))\Big).
\end{align}

An analytical expression is not possible to obtain for the sub-horizon integral. However, on integrating numerically we find it to be negligible compared to the super-horizon integral. For example, if we choose $\sigma = 0.1$, $\eta_V = 0.01$, and $\epsilon=0.015$, then the sub-horizon integral is 24 orders of magnitude smaller than the super-horizon integral. As such we directly make use of the asymptotic form of the Hankel function for $x<<1$(super-horizon) in Eq.~(\ref{phiphi1}) with the error being negligibly small. In that case, using
\begin{align}
    H_\nu^{(1)}(x<<1)\approx\sqrt{\frac{2}{\pi}}e^{-i\frac{\pi}{2}}2^{\nu-\frac{3}{2}}\frac{\Gamma[\nu]}{\Gamma[3/2]}x^{-\nu}
\end{align}
we obtain the two-point function
\begin{align}
    \langle\delta\phi(t)\delta\phi(t')\rangle &= \int d(\text{ln}k)|C_1(k)-C_2(k)|^2\coth{\left(\frac{\beta k}{2}\right)} \ k^{\eta_V}(aH)^{-\eta_V}(a'H)^{-\eta_V}\frac{H^2}{8\pi\sigma^4}\frac{2^{2\nu-5}}{\Gamma(3/2)^2}\frac{2}{\pi}\nonumber\\
    &\times\left[\int_{k\eta}^{k\eta_{i}}dx\left(2x^{\frac{5}{2}-\eta_V-\nu}e^{-\frac{x^2}{2\sigma^2}}\Gamma(\nu)\alpha-\frac{1}{3}x^{\frac{9}{2}-\eta_V-\nu}e^{-\frac{x^2}{2\sigma^2}}\Gamma(\nu-1)\right)\right]^2.
\end{align}
For $t=t'$, the power spectrum reads
\begin{align}
    P_{\delta\phi} &=\frac{H^2}{\pi^3}|C_1(k)-C_2(k)|^2\coth{\left(\frac{\beta k}{2}\right)}2^{2\nu-3}\left(\frac{k}{aH}\right)^{3-2\nu}\sigma^{3-2\nu}\sigma^{-2\eta_V}\nonumber\\
    &\times\Bigg[\Gamma(v)\left(\Gamma\left(\frac{7}{4}-\frac{\eta_V+\nu}{2},\frac{k^2\eta^2}{2\sigma^2}\right)-\Gamma\left(\frac{7}{4}-\frac{\eta_V+\nu}{2},\frac{k^2\eta_{i}^2}{2\sigma^2}\right)\right)\left(1-\dfrac{1}{3}\left(\dfrac{3}{2}-\nu\right)\right)\nonumber\\
    &\hspace{40mm}-\frac{\sigma^2}{3}\Gamma(\nu-1)\left(\Gamma\left(\frac{11}{4}-\frac{\eta_V+\nu}{2},\frac{k^2\eta^2}{2\sigma^2}\right)-\Gamma\left(\frac{11}{4}-\frac{\eta_V+\nu}{2},\frac{k^2\eta_{i}^2}{2\sigma^2}\right)\right)\Bigg]^2. \label{pws1}
\end{align}
Let us define the scale factors $a_R(t)$ and $a_I(t)$ during the radiation-dominated era and the inflationary era respectively. Since we are interested in integrating out modes of wavelength less than or equal to the co-moving Hubble radius at the end of the thermal era, we take the coarse-graining parameter $\sigma$ to be
\begin{align}
    \sigma = \frac{a_RH_R}{a_IH_I}\Big|_{t=t_i} = \frac{H_R(t_i)}{H_I(t_i)},
\end{align}
where we have used $a_R(t_i)=a_I(t_i)$. To keep $\sigma<1$ we expect $H_I(t_i)>H_R(t_i)$.
Let us evaluate $\eta_i^2$ which is given by
\begin{align}
    \eta_i^2 = \frac{1}{a_I^2(t_i)H_I^2(t_i)}\label{etai}.
\end{align}
To find $a_I(t_i) = a_i$ we use
\begin{align}
    a_I(t) = a_ie^{H(t-t_i)}
\end{align}
which gives us
\begin{align}
    a_i = a_I(t)e^{-H(t-t_i)}.
\end{align}
Let us set $t=t_*$ above, where $t_*$ denotes the time when the modes of interest crossed the horizon. We define $k_i = a_iH_i$ to be the wave number of the modes that first left the horizon when inflation started. The modes $k_*$ that have re-entered today were the ones that left after the modes $k_i$ left during inflation. Let us say that the modes $k_i$ and $k_*$ left the horizon $N(k_i)$ and $N(k_*)$ e-foldings before the end of inflation respectively, then defining,
\begin{align}
    \delta N = N(k_i)-N(k_*)
\end{align}
we have
\begin{align}
    a_i = a_I(t_*)e^{-H(t_*-t_i)} = a_I(t_*)e^{-\delta N}.\label{ati}
\end{align}
Now to find $a_I(t_*)$ we note that the modes $k_*$ re-enter today which means
\begin{align}
    k_* = a_I(t_*)H_I(t_*) = a_0H_0 = H_0
\end{align}
giving us
\begin{align}
    a_I(t_*) = \frac{H_0}{H_I(t_*)}.\label{ats}
\end{align}
On using Eq.~(\ref{ats}) and (\ref{ati}) in Eq.~(\ref{etai}) we obtain,
\begin{align}
    \eta_i^2 = \frac{e^{2\delta N}}{H_0^2}\label{etaieval},
\end{align}
which gives us the following expression for the power spectrum (\ref{pws1}).
\begin{align}
    P_{\delta\phi} &=\frac{H^2}{\pi^3}|C_1(k)-C_2(k)|^2\coth{\left(\frac{k}{2T}\right)}2^{2\epsilon-2\eta_V}\left(\frac{k}{k_*}\right)^{2\eta_V-2\epsilon}\sigma^{-2\epsilon}\nonumber\\
    &\times\Bigg[\Gamma\left(\frac{3}{2}-\eta_V+\epsilon\right)\left(\Gamma\left(1-\frac{\epsilon}{2},\frac{k^2}{2\sigma^2k_*^2}\right)-\Gamma\left(1-\frac{\epsilon}{2},\frac{k^2e^{2\delta N}}{2\sigma^2k_*^2}\right)\right)\left(1-\dfrac{1}{3}(\eta_V-\epsilon)\right)\nonumber\\
    &\hspace{40mm}-\frac{\sigma^2}{3}\Gamma\left(\frac{1}{2}-\eta_V+\epsilon\right)\left(\Gamma\left(2-\frac{\epsilon}{2},\frac{k^2}{2\sigma^2k_*^2}\right)-\Gamma\left(2-\frac{\epsilon}{2},\frac{k^2e^{2\delta N}}{2\sigma^2k_*^2}\right)\right)\Bigg]^2, \label{pws}
\end{align}

\begin{figure}[h!]
\centering
  \includegraphics[width=0.7\textwidth]{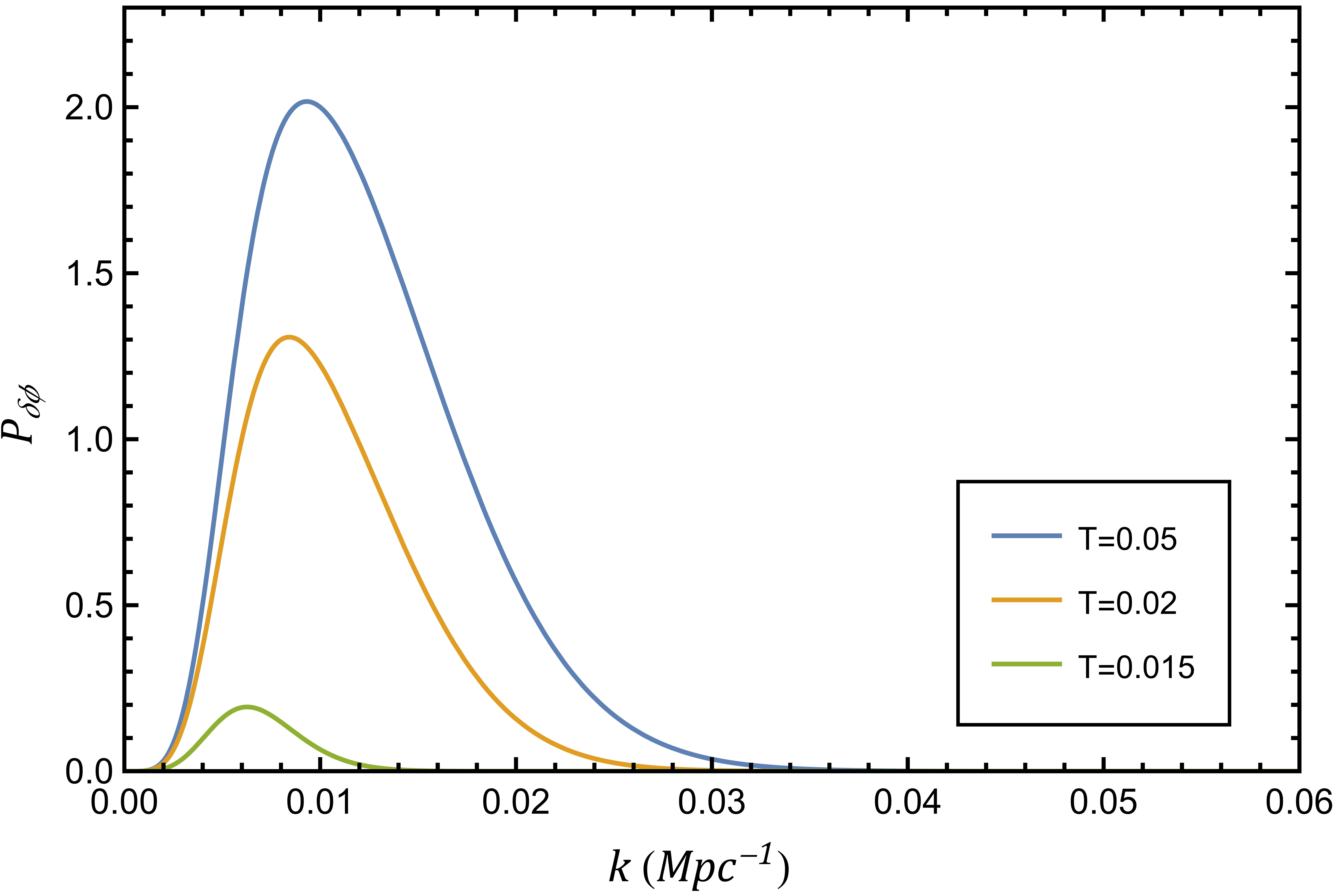}
  \caption{Power spectrum, Eq.~(\ref{pws}) $P_{\delta\phi}$  v/s $k$}\label{Pvsk}
\end{figure}

where~\cite{Das:2014ffa}
\begin{align}
    &|C_1(k)-C_2(k)|^2 = \frac{\pi}{2z_I}\Bigg[\Bigg\{\Big(\nu+\frac{1}{2}\Big)J_{\nu}(z_I)-z_I J_{\nu+1}(z_I)\Bigg\}^2+z_I^2J^2_{\nu}(z_I)\Bigg]\Bigg|_{t=t_i},\\
    &z_I(t_i) = \frac{k}{(1-\epsilon)H_0}e^{\delta N}, \\
    &k_i = a_iH_i= k_*e^{-\delta N}\ \ \ \ \ \text{(when inflation begins)},\\
    &\nu = \frac{3}{2}-\eta_V+\epsilon.
\end{align}
Figure \ref{Pvsk} shows the power spectrum $P_{\delta\phi}$ v/s $k$ in units of $\frac{H^22^{2\epsilon-2\eta_V}}{\pi^3}$ for three different values for temperature with $\sigma = 0.5$.

\section{Statistical analysis and results}\label{sec:analysis}
The Primordial Power Spectrum (PPS) in the stochastic thermal inflation, in addition to the canonical parameters $A_s$ and $n_s$ has 3 new parameters, namely, temperature $T$, separation parameter $\sigma$, 
and the number of e-foldings of inflation in excess of the standard minimum $\delta N$. We implement the modified PPS into the publicly available Boltzmann solver CAMB~\cite{Lewis:1999bs} and analyze using the publicly available Markov Chain Monte Carlo (MCMC) sampler Cobaya~\cite{Torrado:2020dgo}. In addition to the 3 new parameters of the PPS, our Markov Chain Monte Carlo (MCMC) analysis includes the standard 6 cosmological parameters, i.e., the physical densities of baryons ($\Omega_b h^2$) and dark matter ($\Omega_c h^2$), the Hubble parameter ($H_0$), the amplitude ($A_S$) and spectral index ($n_s$) of the primordial power spectrum (at the pivot scale $k_P = 0.05, \text{Mpc}^{-1}$), and the optical depth of reionization ($\tau$). Our prior choices for the new parameters are shown in Table~\ref{tab:priors}. Using these prior ranges, we analyze our model against the Planck-2018-low-$\ell$ (TT), Planck-2018-low-$\ell$ (EE), Planck-2018-high-$\ell$, and  Planck-2018-lensing likelihoods.

\begin{table}[b!]\label{tab:priors}
    \centering
\begin{tabular} {| l|c|}
\hline
 Parameter &  Prior range\\
\hline
\hline
{\boldmath$\log_{10}T   $} & $[-15,0]$\\

{\boldmath$\sigma         $} & $[0,1]$\\

{\boldmath$\delta N       $} &  $[1,15]$\\

{\boldmath$\log_{10}\epsilon$} & $[-8,0]$\\
\hline
\end{tabular}
    \caption{Prior on the three new parameters, $T$, $\sigma$, $\delta N$, and the slow-roll parameter $\epsilon$. All the priors are taken to be uniform in the respective range.}
    \label{tab:my_label}
\end{table}

\begin{figure}[h!]
\centering
  \includegraphics[width=\textwidth]{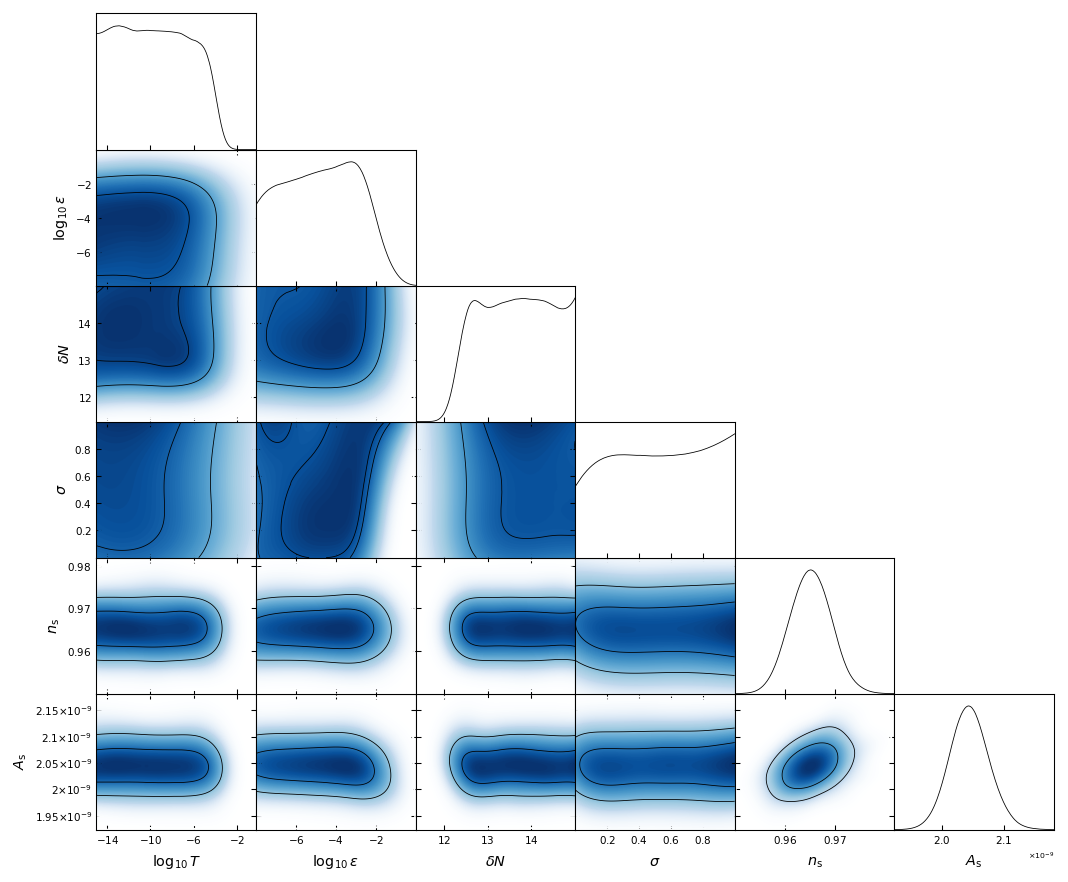}
  \caption{1D and 2D marginalized posteriors for the feature parameters of thermal stochastic inflation, i.e., temperature, $T$, coarse grain parameter $\sigma$, and $\delta N$. In addition, we also kept the slow roll parameter, $\epsilon$, the posterior for which is also shown here. The standard 6 $\Lambda$CDM cosmological parameters were sampled but are not shown here.}\label{plot2}
\end{figure}


\begin{table}[]
    \centering
    \begin{tabular} {| l |  l | l | c |}
\hline
 Parameter &  68\% limits  &  90\% limits &  95\% limits\\
\hline \hline
{\boldmath$\log_{10}T     $} & $-9.5^{+2.2}_{-5.1} $  & $< -4.95  $  & $< -4.31 $\\

{\boldmath$\sigma         $} & $> 0.344 $  & $> 0.161 $ & $ - $\\

{\boldmath$\delta N       $} & $> 13.1 $   & $> 12.5 $ & $> 12.5  $\\

{\boldmath$\log_{10}\epsilon$} & $-4.7^{+2.2}_{-1.9} $  & $-4.7^{+2.5}_{-3.2} $ & $< -1.90 $\\
\hline
\end{tabular}
    \caption{Constraints on the feature parameters of stochastic thermal inflation.}
    \label{tab:my_label}
\end{table}

The energy density contained in the inflaton at the beginning of the inflation is set by an era preceding the inflation, i.e., $\rho_I=\rho_R$. In a radiation-dominated Universe, the energy density evolves as $\rho_R \propto g_{\star} T^{4}$ where $g_{\star}$ is the effective degrees of freedom. In addition, we know that $H_R \propto \rho_R$. The cut-off parameter between the quantum and the classical mode, i.e, the coarse-graining parameter $\sigma$ defined as the ratio of Hubble parameters in the radiation era to that in the inflationary era and hence provides an indirect measure of the number of degrees of freedom in the radiation era. In our analysis, we obtain an upper bound on $\sigma$ at 68\% C.L. which suggests that the radiation era preceding the inflationary era has non-zero degrees of freedom, i.e., the inflation indeed follows a radiation era. However, the present data does not provide enough information to establish more stringent constraints on $\sigma$.
\begin{figure}[t!]
\centering
  \includegraphics[width=\textwidth]{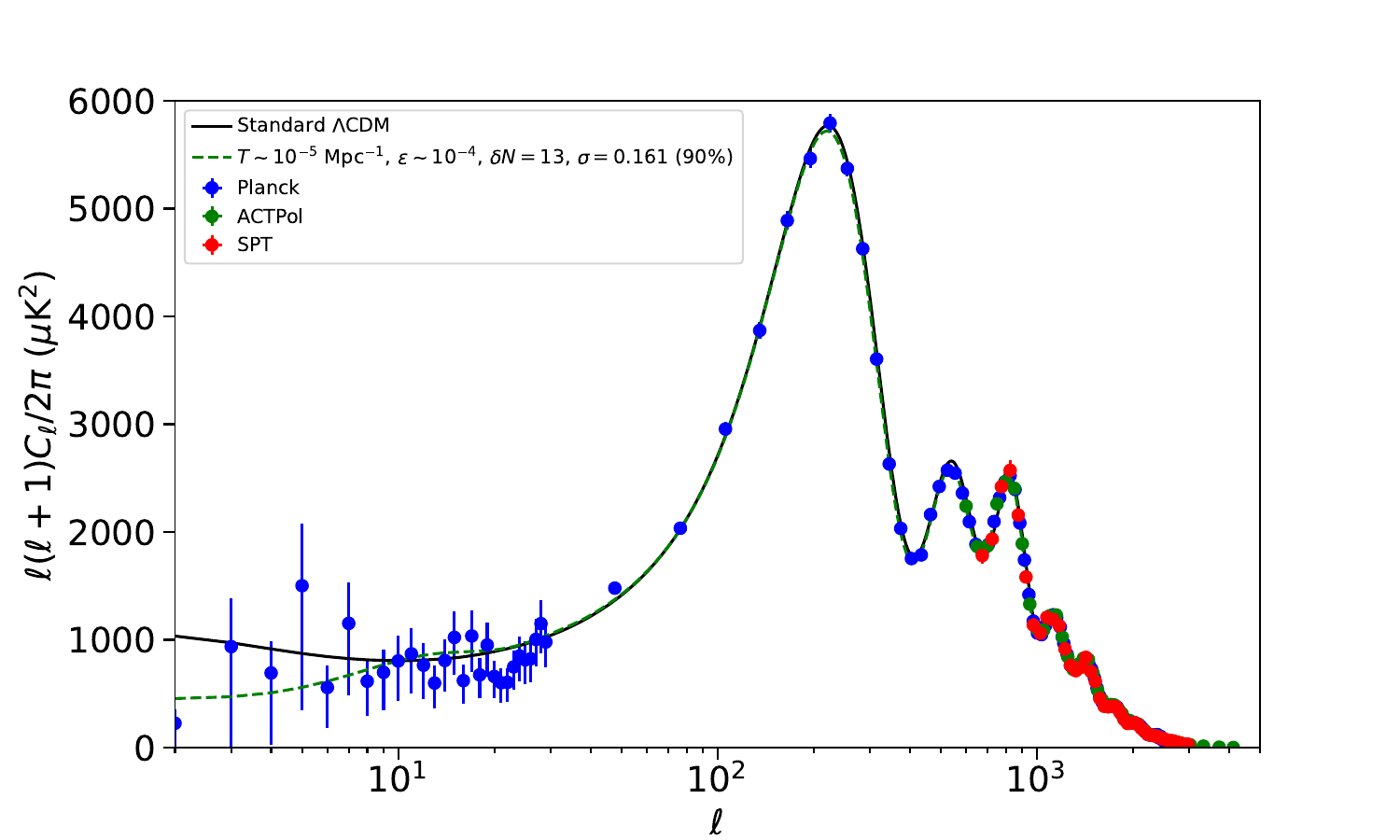}
  \caption{$C_{\ell}^{TT}$ power spectrum, comparison with that for the standard $\Lambda$CDM model along with the observed power spectrum from Planck, ACTPol, and SPT.}\label{plot3}
\end{figure}

\section{Summary and conclusions}
\label{sec:results}
The standard classical inflation explains the observed universe reasonably well. However, it cannot explain the power deficit at the large scale (small multipoles) in the TT power spectrum of CMB. In addition, there is no satisfactory explanation of how quantum modes get converted into classical modes. In this study, we have tried to provide resolutions to the above-mentioned issues by delving into the intricacies of stochastic effects in the thermal inflationary model. Thermal inflation is an intriguing cosmological model in which inflation is preceded by a thermal era, thus setting the initial conditions for inflation. As the thermal era gives way to inflation, the inflationary energy density takes over and the rapid expansion of the universe begins. After $\delta N$ e-foldings, modes that are entering in the present universe, exit the horizon. In classical thermal inflation, the primordial power spectrum is modified by a multiplicative factor, $\coth(k/2T)$, to the standard Bunch-Davies vacuum, where T is the temperature of the inflaton. This results in reducing the power at the small multipoles. However, it still lacks the formalism to explain the transition of modes from quantum to classical. To tackle this issue, we incorporate stochastic effects into the thermal inflation model. Our formalism introduces a coarse-graining parameter, denoted as $\sigma$, which serves as a cutoff scale dividing the modes into quantum and classical. The stochastic thermal inflationary model containing a total of 9 parameters (6 standard and three new parameters $T$, $\delta N$, and $\sigma$) is tested against the cosmological observations from Planck 2018 data release using CAMB Boltzmann solver with the Monte-Carlo sampler Cobaya. 

An important outcome of our analysis against the Planck 2018 data is that the data in fact prefers a non-zero value of the cut-off parameter $\sigma$ hence implying that all but some modes become classical modes. Our model thus provides a much-needed theoretical foundation for the classicalization of quantum modes, shedding light on a long-standing enigma in cosmology.

Furthermore, in our study, we are able to explain the power deficit at large scale (quadrupole modes) in the TT power spectrum of CMB which was reported in multiple CMB observations. However, in the present formalism, the inflaton temperature required to explain the anomaly is orders of magnitude smaller than that reported in studies dealing with the classical treatment of thermal inflation. This adjustment underscores the significance of accounting for stochastic effects and the presence of a cutoff scale in the thermal inflation framework.

In conclusion, our work extends the understanding of thermal inflation by incorporating stochastic effects, offering an explanation for the conversion of quantum modes into classical modes, and updating the resolution of the long-standing power deficit anomaly in thermal inflation by revising the predicted inflationary temperature. These findings enhance our comprehension of the early universe and underline the importance of considering stochastic effects in cosmological models.

\acknowledgments 
The authors acknowledge SAMKHYA: high Performance Computing Facility provided by Institute of Physics, for the performed simulations. AN acknowledges the financial support through APEX project (theory) at Institute of Physics, Bhubaneswar. This work is also partially supported by the DST (Govt. of India) Grant No. SERB/PHY/2021057.

\appendix

\section{Additional Figures}
\begin{figure}[h!]
    \centering
    \includegraphics[width=\textwidth]{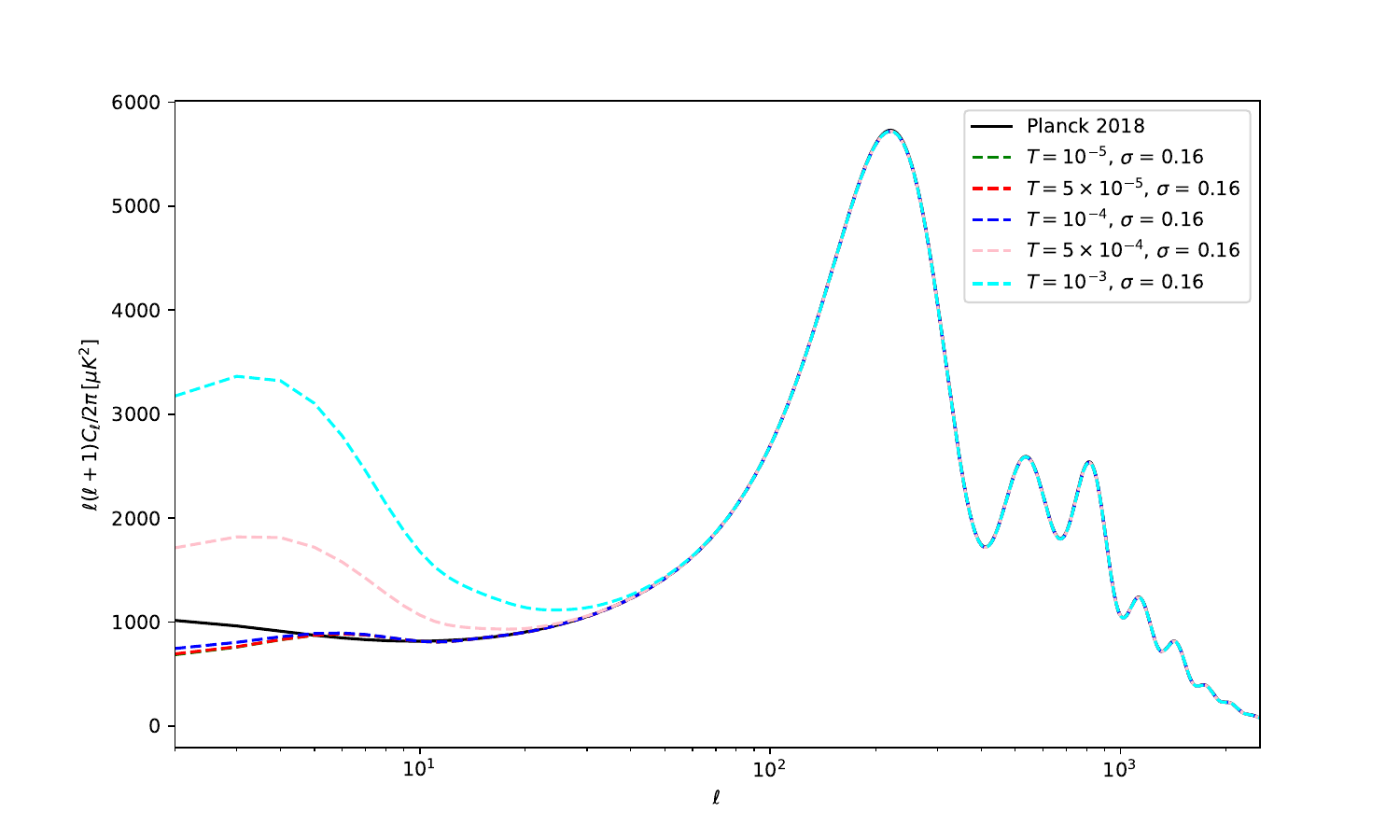}
    \caption{Variation of TT power spectrum with temperature. $\delta N$ fixed at 13.}
    \label{plot4}
\end{figure}

\begin{figure}[h!]
    \centering
    \includegraphics[width=\textwidth]{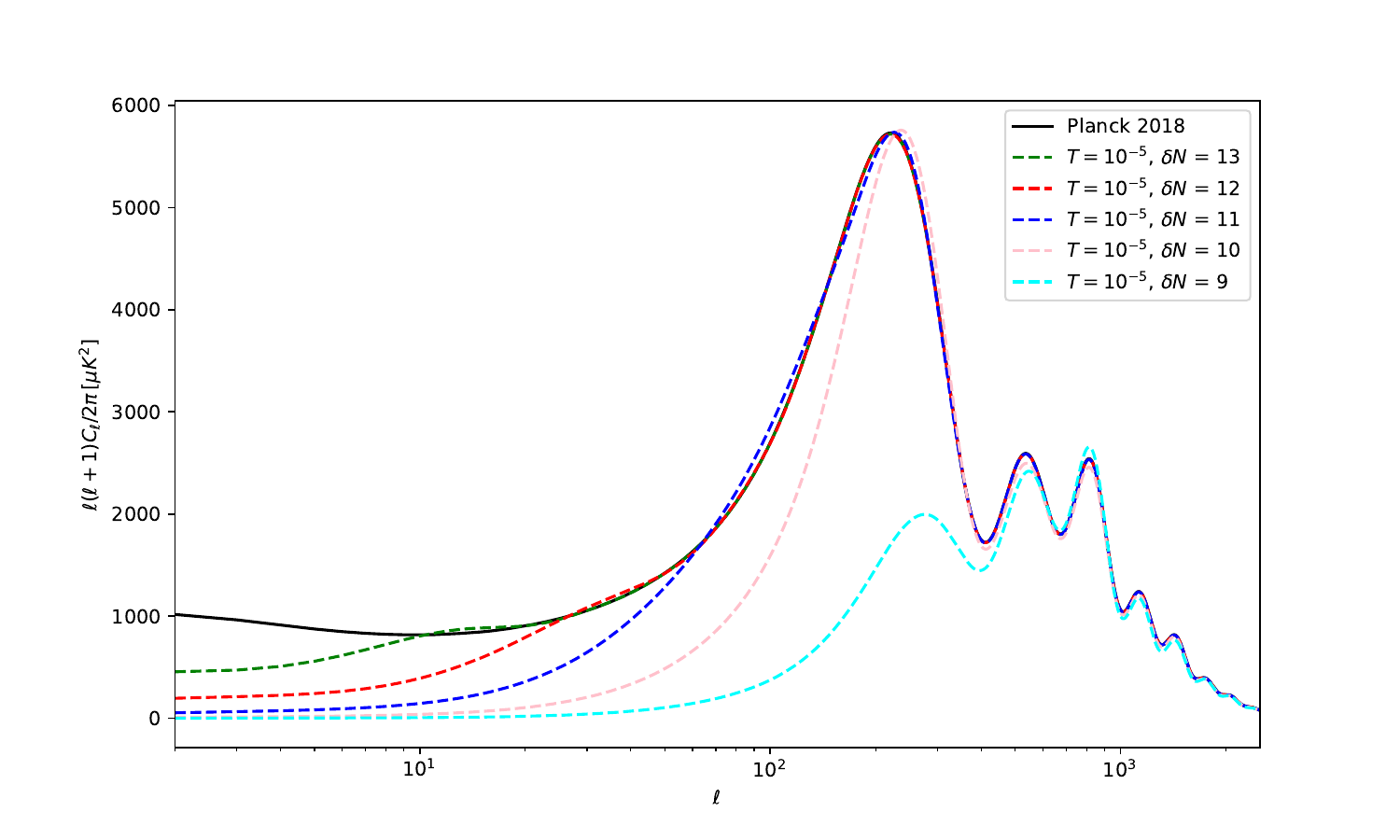}
    \caption{Variation of TT power spectrum with $\delta N$. $\sigma$ fixed at 0.161.}
    \label{plot5}
\end{figure}

\section{Dissipation and mass renormalization}\label{app:dissipation}
\begin{figure}[h!]
\centering
  \includegraphics[width=0.6\textwidth]{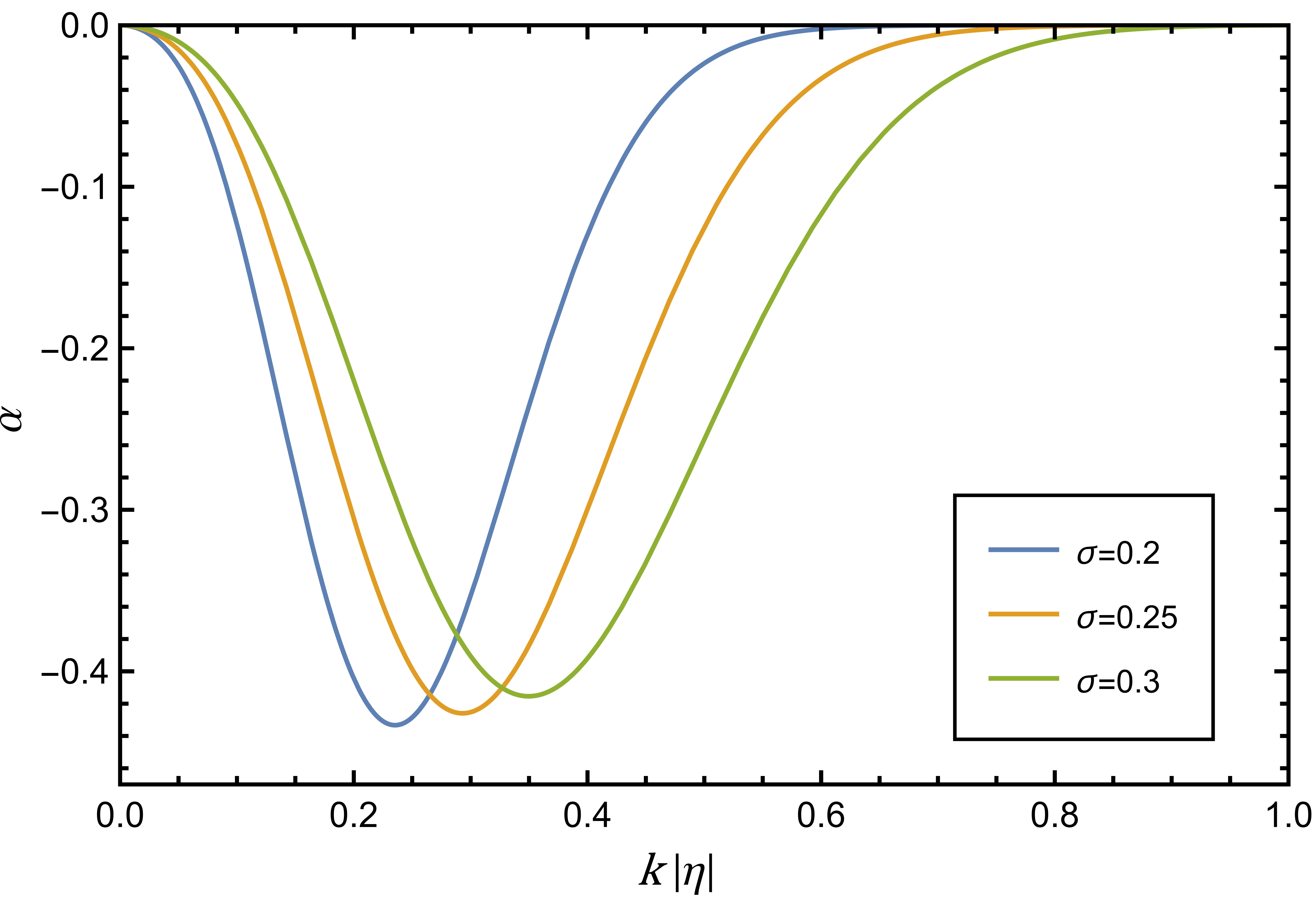}
  \caption{$\alpha$ v/s $k|\eta|$ for three values of $\sigma$. The plots peak roughly at $k|\eta|\sim\sigma$}\label{fig:alpha}
\end{figure}

\begin{figure}[h!]
\centering
  \includegraphics[width=0.6
  \textwidth]{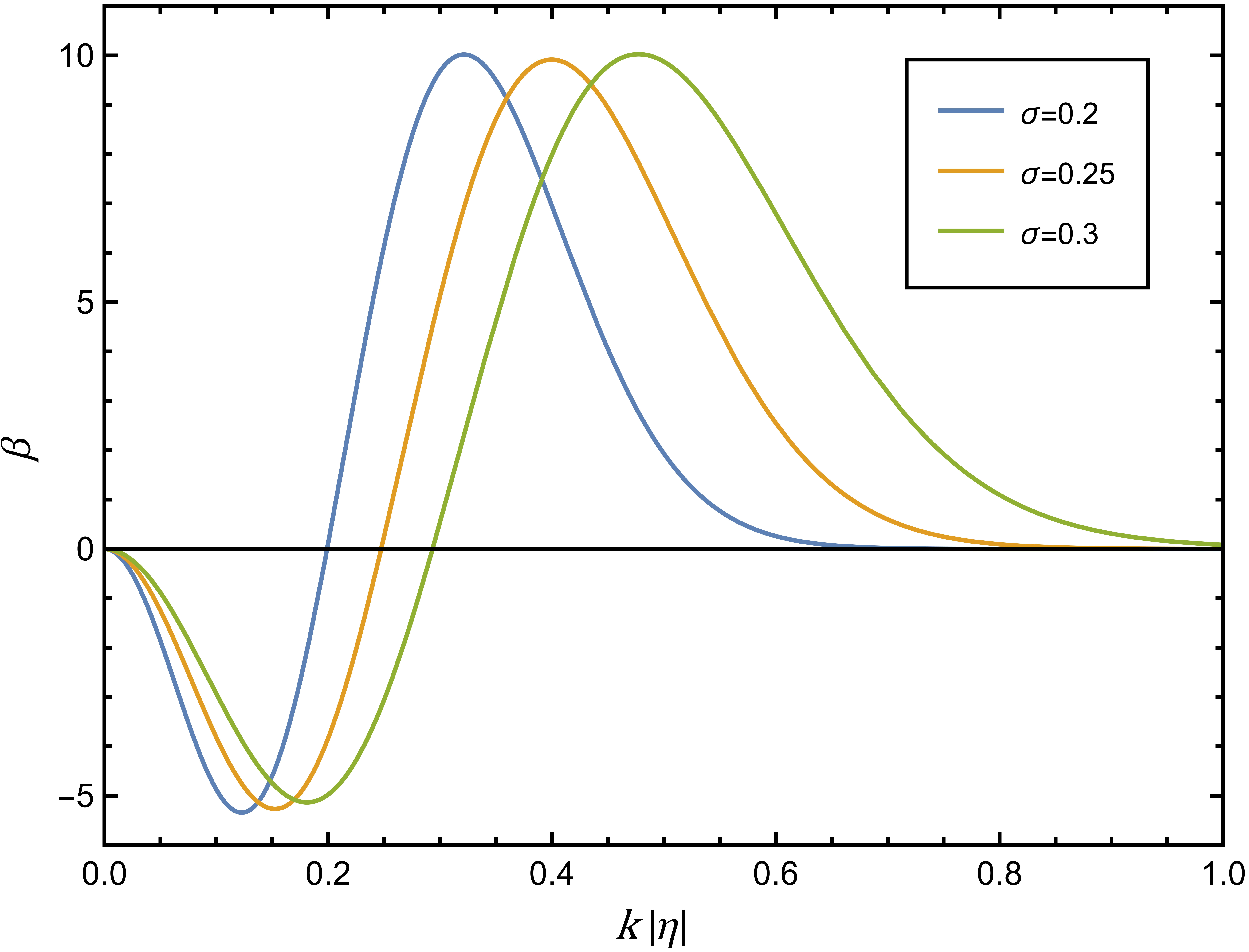}
  \caption{$\beta$ v/s $k|\eta|$ for three values of $\sigma$. We see that the mass renormalization is zero and changes sign at $k|\eta|=\sigma$.}\label{fig:beta}
\end{figure}

The real part of the thermal influence action contributes the following terms to the equations of motion for the scalar field fluctuation.
\begin{align}\label{dissmass}
    \int d^4x' a^3(t')\theta(t-t')\zeta(x.x')\dot\delta\varphi(x')+\int d^4x'a^3(t')\delta m^2(x,x')\delta\varphi(x'),
\end{align}
where
\begin{align}
    &\zeta(x,x') = 2\text{Im}\int d^3\mb{k} \hat P(t)\phi_\mb{k}(x)\dot W(t')\phi_\mb{k}^*(x'),\\
    &\delta m^2(x,x') = -4\delta(t-t')\text{Im}\int d^3\mb{k}\dot W(t)\dot\phi_\mb{k}(x)\dot W(t')\dot\phi_\mb{k}^*(x') \nonumber\\ &\hspace{40mm}- 2\theta(t-t')\text{Im}\int d^3\mb{k} \hat P_t\phi_\mb{k}(x)\dot W(t')\dot\phi_\mb{k}^*(x').
\end{align}
The interpretation of the terms written above is straightforward. The first term is like a friction term similar to $3H\dot\delta\varphi$ and is hence called the dissipation term. The second term goes like the mass term $m^2\delta\varphi$ and is hence called the mass renormalization term. We shall compute these terms and show under what conditions these can be neglected. We start with the dissipation term. For a massive scalar field evolving from an initial thermal state, the dissipation term in Fourier space reads
\begin{align}
    \frac{1}{(2\pi)^{3/2}}\int dt'\int d^3x' a^3(t')\theta(t-t')\left(2\text{Im}\int d^3\mb{k}(\hat P(t) F(k,\eta))\dot W(t')F^*(k,\eta') e^{-i\mb{k}\cdot\mb{x}}\dot\varphi_\mb{k}(t')\right),
\end{align}
where
\begin{align}
    F(k,\eta,\eta') &= \left(\frac{H\sqrt{\pi}|\eta|^{3/2}}{2}\right)\left(C_1(k)H_\nu^{(1)}(k|\eta|)+C_2(k)H_\nu^{(2)}(k|\eta|)\right).
\end{align}
We remind ourselves that $\varphi_\mb{k}$ are the long-wavelength modes. It contains super-horizon modes which vary slowly compared to the sub-horizon modes. As such we take out $\dot\varphi_\mb{k}$ from the integral treating it essentially as a constant, allowing us to express the dissipation term as $\alpha(\eta,\sigma)H\dot\varphi_\mb{k}(t)$, where
\begin{align}
    \alpha(\eta,\sigma) = \frac{2}{H}\text{Im}\left\{(\hat P(t) F(k,\eta))\int_{-\infty}^t dt' a^3(t')\theta(t-t') \dot W(t') F^*(k,\eta')\right\}.
\end{align}
The $\sigma$ dependence of $\alpha$ arises from the window function. Making a change of integration variable $t'\rightarrow x=k|\eta'|$, the integral in the expression for $\alpha(t,\sigma)$ reads
\begin{align}
    \frac{k^{3/2}}{2H^2\sigma^2}\int_{\infty}^{k|\eta|}x^{-1/2}e^{\frac{-x^2}{2\sigma^2}}\left(C_1^*(k)H_\nu^{(1)*}(x)+C_2^*(k)H_\nu^{(2)*}(x)\right).
\end{align}
We split the integral into subhorizon and superhorizon parts as done in Eq.~(\ref{split}). A similar approximation holds in this case too. Following the arguments following Eq.~(\ref{split}), the integral becomes
\begin{align}
    &\frac{k^{3/2}\sqrt{\pi}}{2H^2\sigma^2}\int_1^{k|\eta|} x^{-1/2}e^{-\frac{x^2}{2\sigma^2}}\left(\frac{C^*_1(k)+C_2^*(k)}{\Gamma(\nu+1)}\left(\frac{x}{2}\right)^{\nu}+\frac{i}{\pi}(C_1^*(k)-C_2^*(k))\Gamma(\nu)\left(\frac{x}{2}\right)^{-\nu}\right)\\
    =&Y\left(C^*_1(k)+C^*_2(k)\right) + Z\left(C^*_1(k)-C^*_2(k)\right)\label{diss1},
\end{align}
where
\begin{align}
    &Y = \frac{k^{3/2}\sqrt{\pi}}{2H^2\sigma^2}.\frac{1}{\Gamma(\nu+1)}.\frac{1}{2^{\nu}}\int_1^{k|\eta|}x^{\nu-1/2}e^{-\frac{x^2}{2\sigma^2}}dx,\\
    &Z = i\frac{k^{3/2}}{2H^2\sigma^2}\frac{1}{\sqrt{\pi}}\Gamma(\nu)2^{\nu}\int_1^{k|\eta|}x^{-\nu-1/2}e^{-\frac{x^2}{2\sigma^2}}dx.
\end{align}
The action of operator $\hat P(t)$ on $F(k,\eta)$ in the limit of $k|\eta|<<1$ becomes
\begin{align}
    \hat P(t)F(k,\eta)=HW\left(C_1(k)+C_2(k)\right)+HX\left(C_1(k)-C_2(k)\right),\label{diss2}
\end{align}
where
\begin{align}
    &W =\left(\frac{H^2(k|\eta|)^{\frac{7}{2}}\sqrt{\pi}}{2\sigma^2k^{\frac{3}{2}}}\right)e^{-\frac{k^2\eta^2}{2\sigma^2}}\Bigg[\left(2-2\nu-\frac{k^2\eta^2}{\sigma^2}\right)\frac{1}{\Gamma(\nu+1)}\left(\frac{k|\eta|}{2}\right)^\nu+\frac{2k|\eta|}{\Gamma(\nu)}\left(\frac{k|\eta|}{2}\right)^{\nu-1}\Bigg],\nonumber\\
    &X = -i\left(\frac{H^2(k|\eta|)^{\frac{7}{2}}\sqrt{\pi}}{2\sigma^2k^{\frac{3}{2}}}\right)e^{-\frac{k^2\eta^2}{2\sigma^2}}\Bigg[\left(2-2\nu-\frac{k^2\eta^2}{\sigma^2}\right)\frac{\Gamma(\nu)}{\pi}\left(\frac{k|\eta|}{2}\right)^{-\nu}\nonumber\\&\hspace{80mm}+\frac{2k|\eta|\Gamma(\nu-1)}{\pi}\left(\frac{k|\eta|}{2}\right)^{-\nu+1}\Bigg].
\end{align}
Using Eqs.~(\ref{diss1}) and (\ref{diss2}) the expression for $\alpha$ becomes
\begin{align}
    \frac{2}{H}\text{Im}\left\{XY(C_1^*+C_2^*)(C_1-C_2)+WZ(C_1^*-C_2^*)(C_1+C_2)\right\}.
\end{align}
From the expressions of $C_1(k)$ and $C_2(k)$
\begin{align}
    &C_1(k)=ie^{iz_R}\sqrt{\frac{\pi}{8z_I}}\left[\left(\nu+\frac{1}{2}-iz_I\right)H_\nu^{(2)}(z_I)-z_IH_{\nu+1}^{(2)}(z_I)\right],\nonumber\\
    &C_2(k)=-ie^{iz_R}\sqrt{\frac{\pi}{8z_I}}\left[\left(\nu+\frac{1}{2}-iz_I\right)H_\nu^{(1)}(z_I)-z_IH_{\nu+1}^{(1)}(z_I)\right],
\end{align}
it is clear that
\begin{align}
    \text{Re}\Big((C_1-C_2)(C_1^*+C_2^*\Big) = 1.
\end{align}
Thus we find
\begin{align}
    \alpha(\eta,\sigma) = \frac{2}{H}(XY+WZ).
\end{align}
The plot of $\alpha(\eta,\sigma)$ v/s $k|\eta|$ for various values of $\sigma$ is shown in Figure \ref{fig:alpha}.
The coefficient is always negative \cite{Matarrese:2003ye} indicating an energy gain as modes cross the cutoff $R = (\sigma aH)^{-1}$ to enter the super-horizon field. This "anti-dissipation" is quite small as long as $k|\eta|$ is small compared to $\sigma$.
A similar calculation can be performed for mass renormalization (the second term of \ref{dissmass})  yielding
\begin{align}
   \int d^4x'a^3(t')\delta m^2(x,x')\delta\varphi(x') &= \frac{m^2\delta\varphi}{3\eta_V}\left\{\frac{k^4\eta^4}{\sigma^4}e^{-\frac{k^2\eta^2}{\sigma^2}}\left(2-\frac{k^2\eta^2}{2\nu(\nu-1)}\right) + \Big(MW+NX\Big)\right\}\nonumber\\
   &=\beta m^2\delta\varphi,
\end{align}
where
\begin{align}
    &M = -i\frac{k^{3/2}\Gamma(\nu)2^{\nu-1}}{\sigma^2\sqrt{\pi}}\int_1^{k|\eta|}dx 
 \ e^{-\frac{x^2}{2\sigma^2}}\left\{\left(\frac{3}{2}-\nu\right)x^{-\nu-1/2}+\frac{1}{2(\nu-1)}x^{-\nu+3/2}\right\},\nonumber\\
    &N = -\frac{k^{3/2}\sqrt{\pi}}{\sigma^2\Gamma(\nu)2^{\nu+1}}\left(\frac{3}{2\nu}+2\right)\int_1^{k|\eta|}dx 
 \ x^{\nu-1/2} \ e^{-\frac{x^2}{2\sigma^2}}.
\end{align}
The factor $\beta$ is the ratio of the mass renormalization term and the mass term in the equation of motion. A plot of $\beta$ v/s $k|\eta|$ for various values of $\sigma$ has been given in Figure \ref{fig:beta}. We observe that the mass renormalization vanishes exactly at the cutoff $k|\eta|=\sigma$. We find that the mass renormalization is negligible compared to the mass term in the equation of motion for the perturbations \ref{eq:eom} when $k|\eta|$ is very small compared to $\sigma$.

\clearpage
\bibliographystyle{JHEP}
\bibliography{stochastic}
\end{document}